\documentclass[sigconf]{acmart}

\usepackage{tablefootnote}
\usepackage[subrefformat=parens]{subcaption}
\usepackage{acmart-taps}

\AtBeginDocument{%
  \providecommand\BibTeX{{%
    \normalfont B\kern-0.5em{\scshape i\kern-0.25em b}\kern-0.8em\TeX}}}

\setcopyright{cc}
\copyrightyear{2025}
\acmYear{2025}
\setcctype{by}
\acmConference[CHI '25]{CHI Conference on Human Factors in Computing Systems}{April 26-May 1, 2025}{Yokohama, Japan}
\acmBooktitle{CHI Conference on Human Factors in Computing Systems (CHI '25), April 26-May 1, 2025, Yokohama, Japan}
\acmDOI{10.1145/3706598.3713638}
\acmISBN{979-8-4007-1394-1/25/04}

\begin{document}
\title{Cyberoception: Finding a Painlessly-Measurable New Sense in the Cyberworld Towards Emotion-Awareness in Computing}

\author{Tadashi Okoshi}
 \authornote{These authors contributed equally as first authors.}
\affiliation{
  \institution{Keio University}
  \state{Kanagawa}
  \country{Japan}
}
\author{Zexiong Gao}  
\authornotemark[1]
\affiliation{
  \institution{The University of Tokyo}
  \state{Tokyo}
  \country{Japan}
}
\author{Tan Yi Zhen}
\affiliation{
  \institution{Singapore Management University}
  \country{Singapore}
}
\author{Takumi Karasawa}
\affiliation{
  \institution{Keio University}
  \state{Kanagawa}
  \country{Japan}
}
\author{Takeshi Miki}
\affiliation{
  \institution{University of Cambridge}
  \state{Cambridge}
  \country{UK}
}
\author{Wataru Sasaki}
\affiliation{
  \institution{Nara Institute of Science and Technology}
  \state{Nara}
  \country{Japan}
}
\author{Rajesh K. Balan}
\affiliation{
  \institution{Singapore Management University}
  \country{Singapore}
}

\renewcommand{\shortauthors}{Okoshi and Gao, et al.}

\newcommand{\rev}[1]{{\textcolor{black}{{#1}}}}

\begin{abstract}

In Affective computing, recognizing users' emotions accurately is the basis of affective human–computer interaction. Understanding users' interoception contributes to a better understanding of individually different emotional abilities, which is essential for achieving inter-individually accurate emotion estimation.
\rev{However, existing interoception measurement methods, such as the heart rate discrimination task, have several limitations, including their dependence on a well-controlled laboratory environment and precision apparatus}, making monitoring users' interoception challenging. This study aims to determine other forms of data that can explain users' interoceptive or similar states in their real-world lives and propose a novel hypothetical concept ``cyberoception,'' {\bf a new sense} (1) which has properties similar to interoception in terms of the correlation with other emotion-related abilities, and (2) which can be measured only by the sensors embedded inside commodity smartphone devices in users' daily lives. Results from a 10-day-long in-lab/in-the-wild hybrid experiment reveal a specific cyberoception type ``Turn On'' (users' subjective sensory perception about the frequency of turning-on behavior on their smartphones) significantly related to participants' emotional valence. We anticipate that cyberoception to serve as a fundamental building block for developing more ``emotion-aware'', user-friendly applications and services.

\end{abstract}



\sloppy
\maketitle

\section{Introduction}

\rev{``{\em Interoception} is scientifically defined as the processing of internal bodily stimuli by the nervous system''~\cite{khalsa2018interoception}.}
In simpler terms, interoception is defined as ``the ability to be aware of internal sensations in the body, including heart rate, respiration, hunger, fullness, temperature, and pain, as well as emotion sensations''~\cite{Weir23}. 
\rev{Recent advance in interoception's role in mental health have underscored its significance in conditions such as anxiety, mood disorders, disordered eating, addiction, and somatic symptom-related issues~\cite{khalsa2018interoception}.}
%
In particular, the psychology community is currently actively measuring and using interoception data in their clinical studies, which has proven useful for understanding and improving mental health and other conditions~\cite{khalsa2018interoception}. 

However, measuring interoception is \rev{difficult} in our real-world daily computing lives as it requires access to many low-level bodily functions such as heart rate, respiration, temperature, etc. 
\rev{For example, measuring the heart rate or body temperature requires dedicated physiological sensors, which are not as common as ubiquitous mobile devices such as smartphones. 
Moreover, it is difficult and virtually not possible to conduct heart-beat-based measurement of interoception since existing methodologies (e.g., counting the number of heartbeats) typically require a controlled in-lab environment and are not suitable for the user's real-world live situations with various types of possible noise source for such physiological states.}


\rev{To overcome such limitations, we present the first (to the best of our knowledge) attempt to find the existence of our novel hypothetical concept ``cyberoception.’’ Cyberoception is defined as subjective sensory perception about the 
user’s basic manipulation behavior with their commodity mobile computing devices such as smartphones, being measured through the embedded sensors in such devices, and acts as the same role as interoception in terms of its correlation with emotional ability (particularly the emotional experience in this study).}
In particular, we do not use any data or sensor that is not already being collected by the phone due to the user's regular phone usage patterns -- we did this consciously to ensure that any solution we produced can be easily integrated into existing mobile apps and workflows.

%
More specifically, we focused on the basic operations in computing behavior where people access cyberspace by manipulating their smartphones throughout the day. The basic operations, such as unlocking the phone and turning on the phone, occur in their daily lives so often and so unconsciously that we hypothesized that such operations could be used as a target operation to question the user's subjective perception of the frequency.
If we can discover a similar sense of interception, which can be measured only using the sensors embedded inside our commodity smartphones used daily, the mobile system, along with various types of applications and services, can obtain and utilize such status for their adaptive behaviors toward the realization of ``emotion-aware'' kind services.
We see our cyberoception service as a key building block for many other apps that use the data to improve the health outcomes of mobile phone users.

In the rest of this paper, we first present more background information about the use of interoception data in the psychology community and then provide our design process to identify existing data sources that can be used to infer cyberoception. 
\rev{We conducted a hybrid experiment for 10 days, including a daily-life study and an in-lab emotional psychology experiment with 25 participants. 
Based on the detailed analysis of the data from 22 participants, we examined participants' perceptions of cyberspace activities, interoceptive abilities, and individual emotional traits. 
As the highlight of our findings, we found a correlation between a specific type of cyberoception, ``Turning On.'' and the experience of emotional valence.}

The main contributions of this paper are:
\begin{itemize}
\item \rev{We introduce the novel hypothetical sense of ``cyberoception,'' based on daily smartphone interactions. The concept is defined as having properties similar to interoception, particularly in its correlation with emotional experience.}
\item \rev{We propose a methodology to measure cyberoception using embedded sensors in smartphones without requiring specialized physiological sensors or controlled laboratory environments. This approach enables continuous, non-invasive sensing in real-world settings.}
\item \rev{Through a 10-day hybrid experiment, the study demonstrates for the first time that the specific cyberoception type ``Turning On'' is significantly correlated with participants' emotional valence.}
\end{itemize}

\vspace{-0.3cm}
\section{Background}
This section introduces our research background, namely the concept of interoception and the existing methodologies of measuring interoceptive abilities. 
\rev{Additionally, we highlight the limitations of these traditional methods, paving the way for introducing a novel concept ``Cyberoception'' as an alternative approach.}

\subsection{Interoception}
\textbf{interoception is defined as one's perception, and therefore awareness and understanding, of the physiological state of the body including changes therein~\cite{craig2002you}.} 
In more layman's terms, interoception is defined as ''the ability to be aware of internal sensations in the body, including heart rate, respiration, hunger, fullness, temperature, and pain, as well as emotion sensations''~\cite{Weir23}.


Interoception was first proposed by the British physiologist Charles Sherrington~\cite{sherrington1906integrative}.
Sherrington classified the senses based on the location of their receptors, categorize them into (1) interoception, (2) exteroception, and (3) proprioception.
According to Sherrington, (1) interoception refers to the sense originating from the interoceptive surface, the internal surface of the body, while (2) exteroception is the sensation produced by receptors close to the body surface that are in direct contact with the external environment, including the following senses: sight, hearing, smell, and taste. Furthermore, (3) proprioception is the sensation produced by the movement of one's own body. The sensation of whether a body part is at rest or in motion is (3) proprioceptive sensation.

More recent research \cite{ceunen2016origin} discusses expanding Sherrington's original physiology-bound concept of interoception. Damasio~\cite{DAMASIO_2003} built on Sherrington's viewpoint that interoception forms the foundation of the sense of physical self but proposed that interoception should include proprioception, visceral perception, and the sense of the internal milieu (e.g., temperature and pain), distinguishing it from exteroception. A similar opinion is also supported by Craig~\cite{craig2002you}. \rev{The latest researches use this revised definition~\cite{chen2021emerging}. }
\subsection{Relationship between Interoception and Emotion}
Many existing studies have shown the relationship between one's interoception and other emotional abilities. A study by Stefan Wiens et al.~\cite{wiens} found a positive correlation between interoceptive error in a heartbeat discrimination task and the intensity of categorized emotional experiences. Katkin~\cite{katkin1985blood} reported that people who are accurate at interoception report greater distress in response to noxious stimulation. Barrett's experiment revealed that individuals with greater sensitivity to their heartbeats emphasized feelings of activation and deactivation when reporting their experiences of emotion over time more than those with lower sensitivity.'' \cite{barrett2004interoceptive}.  

Moreover, the relationship between the interoceptive ability and the ability to recognize and respond to {\textbf other person's emotion} is confirmed. 
A study by Terasawa et al. showed a positive correlation between performance on a heartbeat counting task and the ability to recognize facial expressions of others ~\cite{terasawa2014interoceptive, terasawa2015attenuated}, suggesting a correlation between the interoceptive ability and the ability to recognize emotional experiences.
Georgiou et al. reported that heartbeat-sensitive individuals recognize others’ facial expressions of sadness and fear better than individuals who are less sensitive~\cite{georgiou2018see}. 
Imafuku et al. revealed that people with good interoception act more spontaneous facial mimicry than people with poor interoception~\cite{imafuku2020interoception}. 


\subsection{Measurement of Interoception}
\label{sec:MeasurementOfInteroception}
Thus far, several experimental methodologies related to the state of human internal organs have been proposed to measure the interoception ability.
Interoception is strongly related to the autonomic nervous system, which mainly unconsciously maintains homeostasis by the antagonistic function of the sympathetic and parasympathetic nervous systems.
Thus, the various states of internal organs are controlled by these sympathetic and parasympathetic nerves.

The heartbeat counting is the most well-known method for measuring interoception. 
The ``heart rate counting task'' proposed by Schandry in 1981~\cite{schandry1981heart}, in which the heart rate data ($N_{real}$) from an electrocardiogram (ECG) or pulse sensor is used as the ground truth. Simultaneously, the participant is asked to report their subjective answer on the heartbeat count ($N_{report}$) and not to touch any part of their body. The calculated error rate between $N_{report}$ against $N_{real}$, shown in Equation~\ref{eq:IError}, is defined as ``Interoceptive Error'' and is often used as a metric to evaluate how accurate participant can count their heart rate. Since participants are required to provide subjective responses without physically touching their bodies, the Interoceptive Error effectively reflects their internal bodily awareness rather than relying on haptic sensations.

\begin{equation}
Interoceptive Error =  \frac{|N_{real}-N_{report}|}{N_{real}}
\label{eq:IError}
\end{equation}
\vspace{0.5cm}

The visceral sensation is another well-known approach for measuring interoception. 
The visceral sensation is a type of sensation inside the body and is considered a typical interoception~\cite{ceunen2016origin}. Many methodologies were proposed to measure the interoception ability by measuring the accuracy of sensation to different viscera, including but not limited to gastric signals~\cite{van2016water,smith2021gut}, colon~\cite{whitehead1990tolerance} and bladder~\cite{jarrahi2015differential}. 

\subsection{\rev{Limitation in Existing Methodologies of Interoception Measurement}}
\rev{However, the measurement of interoception statuses introduced above usually requires a well-controlled laboratory environment and physiological sensors since it is mainly based on a behavioral experiment of cardiac perception.} 

\rev{For example, in the heart rate counting task introduced in the previous section, the participant's gesture (not to touch the body during counting), precise ECG sensor, and soundproof environment need to be appropriately controlled to achieve accurate measurement. 
Moreover, the instructor must be well-trained to explain the procedure clearly and in detail to ensure the participant can understand the procedure and do as required.
Furthermore, even with the support from the instructor and the use of professional-grade accurate devices, we cannot avoid collecting some noises from the ECG sensor. }

\rev{In contrast, daily computing involving various information services, such as social networking platforms, occurs not in well-controlled laboratory settings but in dynamic, ``wild'' environments. This computing takes place continuously throughout the day, from morning to night, amidst the noise and complexity of real-world conditions.}

\rev{Despite the importance of measuring interoception status to understand better the users' emotion-related abilities (e.g., how well the users can estimate each other's emotions), the existing methodologies do not match such computing environment in our real lives. In other words, when we aim to construct a system that measures the user's interoceptive ability continuously and repeatedly in such an environment, the aforementioned requirements of the existing methodologies are significant hurdles to the actual implementation.}

\section{Related Work}

\subsection{Measurement Error in Self-Reported Smartphone Usage}
Several studies have highlighted a gap in users' self-reported smartphone usage and actual smartphone usage. This discrepancy inspired the experiment design of cyberoception, which aims to measure users' unconscious intuition about their smartphone habits.

Focusing on communication-oriented smartphone usage, a previous study compared users' self-reports on the frequency of voice calls, SMS, and Gmail with actual logs~\cite{kobayashi2012no}. The users were found to subjectively overestimate the frequency of such voice calls, SMS, and Gmail. Although factors that contribute to overestimation remain unclear, this study argues that such measurement error is not random. However, multiple studies have shown that screen time tends to be subjectively underestimated ~\cite{lin2015time, douglas2012digital, felisoni2018cell}. The mixed results of overestimation and underestimation suggest that we should analyze the various smartphone usage types separately rather than treating them uniformly. 

Several studies have endeavored to identify demographic factors that affect the individual measurement error of self-report in smartphone usage. However, the effects of gender, age, marital/non-marital status, job, and educational status were found to be insignificant~\cite{shum2011evaluation,boase2013measuring}. One critical oversight in these studies is the potential role of affective states in shaping individual perceptions of smartphone use.

Previous studies have collected self-reported data on smartphone usage at weekly or daily intervals. 
For our experiment, we decided to collect self-reports on smartphone usage every 30 minutes to estimate their intuition better.

\subsection{Estimating Mental and Physical States Using Smartphone Data}
The ubiquitous nature of smartphones offers the opportunity to gain valuable insights into people's mental and physical states using data collected from these devices. Numerous studies have mostly used data collected from smartphones to estimate users' emotions, mood, engagement, and physical health. Our study leverages smartphone data to ewxplore interoception, which is related to but different from concepts such as emotion and mood. 

An early work by LiKamwa~\cite{likamwa2011can} showed the feasibility of mood inference from patterns in application usage, phone calls, SMSs, emails, web browsing history, and location. This was expanded with a system called ``MoodScope'' which linked the self-reported mood of the user with their smartphone usage patterns. ~\cite{likamwa2013moodscope} Beyond basic emotions, ``MoodExplorer'' also examined smartphone usage patterns and sensing data to automatically detect compound emotion, defined as a combination of different basic emotions. ~\cite{zhang2018moodexplorer}
In addition, engagement with a task can also be inferred through the user's expression using acoustic sensing from a smartphone. ~\cite{kar2023expressense} Other than estimating mental states, numerous studies have also measured physical health indicators such as heart rate variability and blood pressure using data from a smartphone's built-in accelerometer and camera ~\cite{huynh2019vitamon, wang2018seismo}

There is growing evidence that the various data generated with smartphones can effectively reflect users' mental and physical states. As such, our study aims to explore the potential of using people's perception of smartphone usage can be used to assess interception.

\subsection{\rev{Measurement of Interoception on Smartphones}}
\rev{There is a pre-existing literature on measuring interoception using a smartphone, smartphone-based phase adjustment task (PAT) by Plans~\cite{PLANS2021108171}. PAT is a novel smartphone-based method for assessing interoception by requiring participants to synchronize auditory tones with their heartbeats. In this task, tones are presented at a heart rate-matched frequency but deliberately out of phase with the heartbeats, and participants adjust the phase until they perceive synchrony. 
The method is robust against physiological or strategic confounds and demonstrates variability in performance across individuals.}

\rev{While PAT provides a novel method of measuring interception on the phone, it still relies on actual heartbeat data of the user sensed through a physiological sensor. Meanwhile, our research approach does not focus on directly measuring interception. Rather, we aim to explore new senses that can be measured through non-physiological sensors.}

%

\section{Cyberoception}
To overcome the aforementioned limitation in existing methodologies for measuring interception, we propose our novel hypothetical concept of  ``cyberoception'' and investigate its possibility.

\subsection{Cyberoception: Our Proposing Concept}
\rev{We define our hypothetical concept of ``cyberoception'' as follows.}

\begin{quote}
\rev{Cyberoception is a sense in humans related to their almost unconscious manipulation of a computing device, and can be measured through embedded sensors in such devices without relying on physiological sensors. It serves the same role as interoception in terms of its correlation with emotional ability.}
\end{quote}

\rev{The following subsections will explain, step by step, the background that led to this concept.}

\subsubsection{\rev{Use of Embedded Sensors in Mobile Devices}}
\rev{As described in Section 2, in the user's real-world daily computing situations that we are targeting in this research, the existing interoception measurement methodologies with physiological sensing have limitations. 
Instead, we can see newer opportunities for using embedded sensors in users' mobile devices such as smartphones. 
If we could use such sensors for measuring interoception, that would be a significant step forward in enabling continuous, non-invasive emotion-aware services in real-world settings, leveraging widely accessible mobile devices.}
 
\subsubsection{\rev{Focusing on User's Sensory Perception of Device Interaction}}
\rev{
But at the same time, we need to carefully consider ``what to measure'' by using such embedded sensors. 
If we were to strictly adhere to measuring original interoceptive senses, we would ultimately need to measure internal bodily states, such as heart rate, by some means, thereby reverting the discussion to its initial scope.
}

\rev{On the other hand, looking at real-world computing in our daily life with widespread commodity mobile devices, we see our continuous interaction with these devices throughout our daily lives from morning to night. 
Heavy phone users carry and use their smartphones literally ``always'' throughout their lives, manipulating the phone interfaces and experiencing their interaction with the cyber world.
}

\rev{Regarding this highly continuous and frequent smartphone usage, we consider the possibility that these operational manipulations have become, to some extent, ``unconscious behavior'' for us. (For instance, people may be unaware of how often they lock/unlock their smartphones in a single day, as this behavior is performed so frequently and unconsciously.) }

\rev{Thus, when we think about their subjective ``sense'' around 
such almost-unconscious manipulation, we consider if such sense may have some similarity with the existing sense of interoception in terms of the correlation with the user's ability to regulate the emotion.}






\subsubsection{\rev{Properties Similar to Those of Interoception}}
\rev{As an additional idea, from a different perspective, we hypothesize that devices like smartphones, which serve as interfaces connecting us to cyberspace, could be seen as extensions of our bodies.}

\rev{As presented in Section 2, recent research~\cite{ceunen2016origin,DAMASIO_2003,craig2002you} discuss expanding Sherrington's original physiology-bound concept of interoception and suggests that interoception should encompass proprioception, visceral sensation, and the perception of the internal milieu. 
The evolution of this series of discussions inspired us to discuss whether a further conceptual extension of interoception (including proprioception) was possible even to the cyber world.}

\rev{As we review the related literature, we see that the flexibility and extendability of body schema have been extensively demonstrated in studies on the concept of the extended self~\cite{heersmink2020varieties}. In this context, various tools and objects, including smartphones, have been investigated as extensions of the ``self'', with findings indicating effects comparable to those observed in the ``rubber hand illusion.''~\cite{liepelt2017self} 
Notably, psychological research has consistently shown that smartphones (not limited to their usage but their existence itself) are perceived as an extension of the self~\cite{park2019smartphone,gertz2021smartphone}. 
Moreover, in recent research, interoception is revealed to be the core element of the cognitive process of self~\cite{quigley2021functions}.
}


\rev{Based on this discussion, we have reached the idea that we could possibly hypothesize that cyberoception can be treated as one type of interoception and that it can be inferred to share the emotion-related properties~\cite{DAMASIO_2003,vaitl1996interoception} similar to interoception.
More concretely, since we are focusing on interoception's specific property in terms of the correlation against other emotional abilities (e.g., the ability to recognize and respond to other persons' emotions), we hypothesize that cyberoception holds such similar property. 
}




\subsection{Research Questions}
As the very first step of our interoception research, 
the research questions in this paper are the following two.

\begin{itemize}
\item \textbf{RQ1: Does cyberoception have similar \rev{emotion-related} property to interoception?} : Interoception has long been studied in the fields of psychology and physiology. Various important relationships have been observed, including one's emotion formation, emotion recognition, and estimation of others' emotions. 
In this study, we focus on emotional experience and investigate 
whether the emotional experience of each individual is correlated to his/her cyberoception. We also discuss whether it has properties similar to those of interoception. 

%
\item \textbf{RQ2: Can cyberoception replace \rev{the existing physiological sensor-based measurement methodology of} interoception?} : 
\rev{Assuming that cyberoception has similar emotion-related properties to interoception, additionally, we want to know if cyberoception can be an alternative smartphone-sensor-based measurement methodology of pre-existing interoception which until now has required physiological sensors and a strictly controlled environment.} 
We investigate the relationship between interoception and various types of candidates of cyberoception by comparing interoception measured using a heartbeat counting task with our proposed cyberoception.
\end{itemize}


\section{Measurement of Cyberoception}
\rev{Cyberoception and interoception share a fundamental goal: to capture the user's internal states, although through different means. Interoception gathers physiological signals that are indicative of emotional and cognitive states. However, cyberoception leverages non-physiological data derived from daily interactions with mobile devices to estimate similar internal states. The measurement of cyberoception in this study is inspired by the typical gauge of interoception.}

According to the basic approach of cyberoception, which relies on the (non-physiological) available data on the user's mobile devices, the basic approach to estimating the user's states is based on the collection of (1) the data related to the user's device operation, along with (2) the user's subjective sense on such an operation.

What specific sensations related to smartphones should we measure? After an extensive discussion among the research team, we selected the following six types of operations based on the following criteria: (1) they are operations and actions that users perform daily on their smartphones, and (2) they are extremely daily operations and actions, there is a considerable possibility that they are performed almost unconsciously.

\rev{interoception is closely tied to the autonomic nervous system, encompassing sensations such as heartbeat, respiration, and gastrointestinal activity, which are processed almost unconsciously. In proposing Cyberoception as a concept analogous to interoception, it is crucial to account for this characteristic of almost unconscious processing inherent in interoceptive sensations. Therefore, we adopted the criterion that operations and actions selected for Cyberoception metrics should be those that are extremely daily and thus likely to be performed almost unconsciously. Pragmatically, focusing on extremely daily operations makes Cyberoception measurable and feasible to study. These frequent and routine actions provide a practical foundation for collecting consistent data, enabling researchers to capture patterns of user behavior that reflect internal states without requiring invasive or burdensome methods.}

The selected six types are  (1) Turning On, (2) Unlocking, (3) Screen Use Duration, (4) Micro-usage, 
(5) Most-used App and (6) Typo (Typographical Error). 
We call each of these six types the candidates of \textbf{Cyberoception Metric} for the rest of the paper.

%
%

\subsection{Turning On}
{\textbf Turning On} represents a user's activity to turn on the smartphone screen.
Turning On represents the user intentionally switching the smartphone screen on for full user interaction \rev{(not ambient display or other non-interactive state)} and is a good indicator of the frequency of smartphone use. \rev{Turning On activity hypothetically may reflect two types of emotional states: (1)aroused and motivated affect with the attention towards to smartphone  (2)acquiring a sense of reassurance or calm by escaping negative emotions based on the attachment theory.}

On the concrete methodology of collecting Turning On activity data, they can be sensed by an API on the smartphone OS, such as {\textit android.app.usage API}, which records the type of usage activity and timestamp.

For clarification, (non-manual) automatic activation of the phone display to show the time and/or weather during the sleep mode is excluded from the definition of Turning On since it does not represent the user's intentional switching on activity.

\subsection{Unlocking}
{\textbf Unlocking} measures the frequency of a user unlocking the phone.
Unlocking requires the user 
to intentionally enter the PIN, face, or fingerprint authentication to unlock the screen.

For the data collection methodology, Unlocking activity can be sensed by an API on the smartphone OS, such as by monitoring keyguard hidden activity in the Android platform. 

Turning On and Unlocking are different operations, although they are related. Unlocking often occurs immediately after Turning On when the user turns on and unlocks the phone in one sequence. Thus, the number of times the screen is turned on is significantly related to the number of times the screen is unlocked\rev{, and the emotion accompanied by Unlocking should be similar to Turning On}. Despite this, there are some cases where a user turns on the screen, checks the time, and turns off the screen without unlocking it. \rev{Compared to Turning On, the Unlocking operation should be affectively more conscious.}

We hypothesize that the users initiate such Turning On and Unlocking operations so frequently and almost unconsciously ~\cite{wilcockson2018determining} in their daily lives that the measurement of such operations can be good candidates for the cyberoceptive measurement methodology. 

\subsection{Screen Use Duration}
{\textbf Screen Use Duration} is the user's cumulative use time of the smartphone screen within one session.
In a typical case, this is the duration that starts from turning-on the phone, unlocking, one or multiple application usages, and finishing with locking and/or turning-off the phone.

The APIs provided by the smartphone OSs and used to record \textbf{Unlocking} and \textbf{Turning On} can be combinedly used for the calculation of \textit{Screen Use Duration}.

\textbf{Screen Use Duration} is a metric often considered when using smartphones. The perception of screen time could be partly determined by time perception ability. But more importantly, it should be related to their sense of smartphone use. \rev{The discrepancy between self-reported Screen Use Duration and objective usage data may suggest signs of smartphone addiction. Emotions commonly associated with smartphone addiction, such as guilt, frustration, or anxiety, could potentially play a role in this context.}

\subsection{Micro-Usage}
\textbf{Micro-usage} is defined as a user's application usage session under 5 seconds.
The usage sessions that end within 5 seconds of the user's application launch are recorded as \textbf{micro-usage}.

For Micro-usage data collection, we can leverage the APIs provided by the smartphone OSs, for instance, UsageEvents.Event on 
Android. 
Via such APIs, the user's application launching and ending activities can be recorded continuously. If the duration of the session specified by launching and ending activities is less than 5 s, such usage sessions are marked as Micro-usage.

Research on smartphone usage trends has focused on sessions with short usage times (Micro-Usage), particularly regarding application usage time.
In an extensive survey of 4125 people in 2011, it was found that Android smartphone users surprisingly used applications for less than 5 seconds in 49.9\% of their sessions~\cite{bohmer2011falling}. The average duration of sessions with social networking applications is short~\cite{ferreira2014contextual}, accounting for most of the sessions less than 5 s. However, it is infeasible to interpret all the sessions used for less than 5 s with only social networking app use.

In a study of 21 subjects, Ferreira et al. used 15 seconds as a boundary line in their analysis since nearly 50\% of the collected sessions were less than 15 seconds long. However, Shepard et al. study of 25 subjects \cite{shepard2011livelab, yan2012fast} and Andrews et al. study of 23 subjects~\cite{andrews2015beyond} found that 30-second sessions accounted for half of the sessions. The distribution of smartphone use sessions was influenced by the bias of the subjects selected for the experiment. Natureally, the general trend of smartphone use at the time of the experiment changes constantly and that experiments conducted over different periods lead to different conclusions.
furthermore, usage of less than 15 s was claimed as habitual checking behavior, and the frequency of such usage was proposed as an indicator to evaluate smartphone addiction~\cite{andrews2015beyond, wilcockson2018determining}. \rev{Such habitual checking behavior of Micro-usage hypothetically indicates emotion states related to smartphone addiction. }

\subsection{Most-used App}
The Most-used App represents the launching activity of the user's self-reported most frequently used application.
Specifically, the user's launching operation and move-to-foreground operation of the most used application are counted as the Most-used App.

Most-used App can be sensed when android.app.Activities is moved to the foreground.

The application launch includes multiple launches within a single screen turn-on. Therefore, the sensation of application launches is a more sensitive sensation. However, it is challenging to capture the number of times all applications launch, and it can be assumed that most participants cannot capture the number of times all applications launch correctly. Therefore, in the demographic survey before the start of the experiment, we obtained the names of the ``most frequently used applications,'' and measured the participants' perception of the application's launch. \rev{The emotion accompanied by ``Most-used App'' is hypothetically similar to ``Turn On'' but in a more micro scope.}

\subsection{Typo (Typographical Error)}
Typo (Typographical Error) is a typical fundamental operation in smartphone use, representing a mistake made while typing on a smartphone keyboard.
Generally, the user will correct Typo by pressing the backspace key or using the auto-correction feature.

Many studies have estimated emotion from typing patterns, including mistyping on smartphones, and the number of backspace key presses (delete key) was entered into an emotion estimation model as representative of mistyping~\cite{ghosh2017tapsense, ghosh2019representation, hari2022affectpro}. The number of typographical error recorded by pressing the backspace key was a good representation of the user's emotional state.
In this study, we sensed backspace key presses as the representation of Typo activity in participants' daily lives. \rev{However, auto-correction and predictive text features may affect the accuracy of the typographical error detection. To obtain data that closely reflects users' daily smartphone usage, we refrained from giving any instructions that could alter their daily typing habits. } 
Furthermore, participants' Typo activities were detected more precisely \rev{though non-daily} in-lab experiment.

In this study, we developed an application that counts and records the number of presses of the delete key, and collected data on the number of Typo (Typographical Errors).

\section{System and Experiment Design}
This section details our design of the experiment along with the original system design developed for this experiment. 
After we introduce the overview of the experiment procedure in Section~\ref{sec:OverviewOfExperimentProcedure}, 
we describe the details of 10-day-long data collection/survey study in 
Section~\ref{sec:DailyCeberoceptionMeasuringSurvey}, 
and finally, we present the detailed design of three in-lab experiments in 
Section~\ref{sec:interoceptionexperiment}, \ref{sec:affectivepicture}, and \ref{sec:typingexperiment}).

\subsection{Overview of Experiment Procedure}
\label{sec:OverviewOfExperimentProcedure}

\begin{figure}[t]
    \centering
    \includegraphics[width=1.0\linewidth]{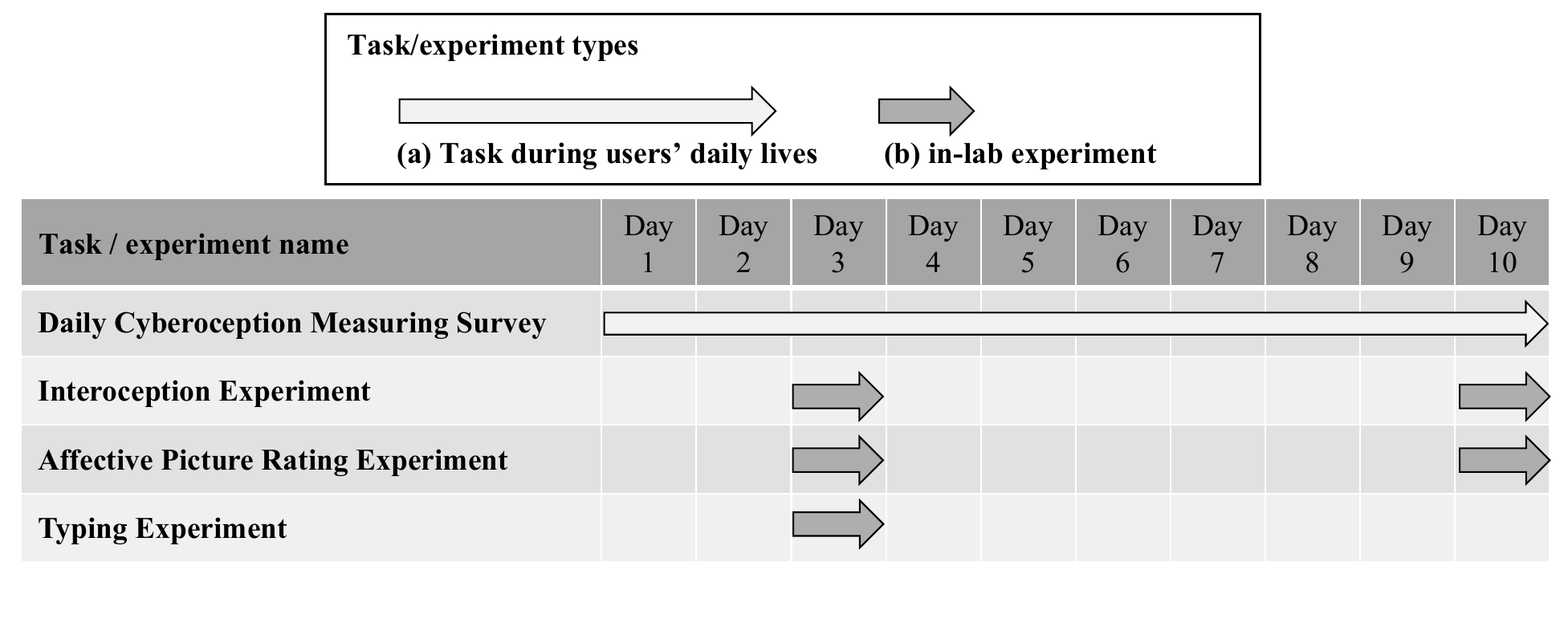}
    \caption{Tasks and Schedule of the study}
    \Description{Tasks and Schedule of the study. It shows the overview of this study containing both (a) a 10-day long data collection/survey period in the user's daily life environment and (b) in-lab experiments on Day 3 and Day 10.} 
    \label{fig:Schedule}
\end{figure}

To measure participants' cyberoception represented by the perception of various basic smartphone operations in daily use and to clarify the relationship between participants' daily cyberoception and their psychological traits, we designed a 10-day experiment.
Figure~\ref{fig:Schedule} shows the overview of this study containing both (a) a 10-day long data collection/survey period in the user's daily life environment and (b) in-lab experiments on Day 3 and Day 10. The (a) task and (b) experiment
are designed separately but analyzed together after the entire survey. 

Throughout the 10-day study period, we tracked participants' daily cyberoception. During these ten days, periodically ``\textbf{Daily Cyberoception Measuing Survey}'' (along with smartphone data collection) (Section \ref{sec:DailyCeberoceptionMeasuringSurvey}) is conducted, where the smartphone usage data and participants' subjective cyberoception questionnaire responses are collected continuously from each participant's smartphone.

On Day 3, three types of in-lab experiments, namely \textbf{Interoception Experiment} (Section~\ref{sec:interoceptionexperiment}), \textbf{Affective Picture Rating Experiment} (Section~\ref{sec:affectivepicture}), and \textbf{Typing Experiment} (Section~\ref{sec:typingexperiment}) are conducted in our laboratory environment. \textbf{Interoception Experiment} is conducted to accurately measure the interoceptive ability of the participant. \textbf{Affective Picture Rating Experiment} involves evaluating the participant's individual traits of emotional experiences. \textbf{Typing Experiment} involves acquiring the participant's sense of Typographical Error in an in-lab environment.

\begin{figure}[h]
    \begin{minipage}[h]{0.45\linewidth}
    \centering
    \includegraphics[width=\linewidth]{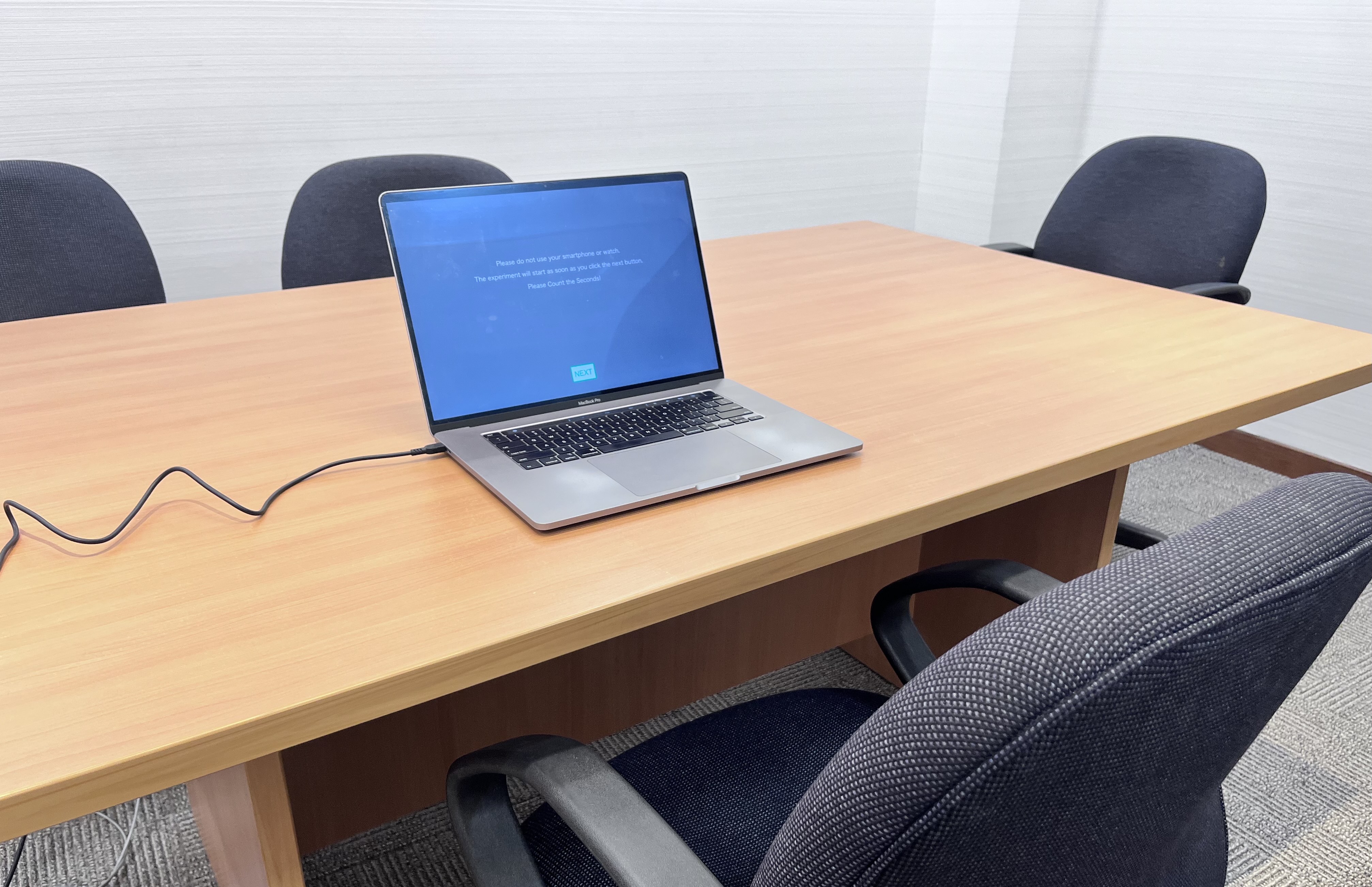}
    \subcaption{Participants were seated in front of a notebook PC.}
    \label{environmentOuter}
    \end{minipage}
    \begin{minipage}[h]{0.45\linewidth}
    \centering
    \includegraphics[width=\linewidth]{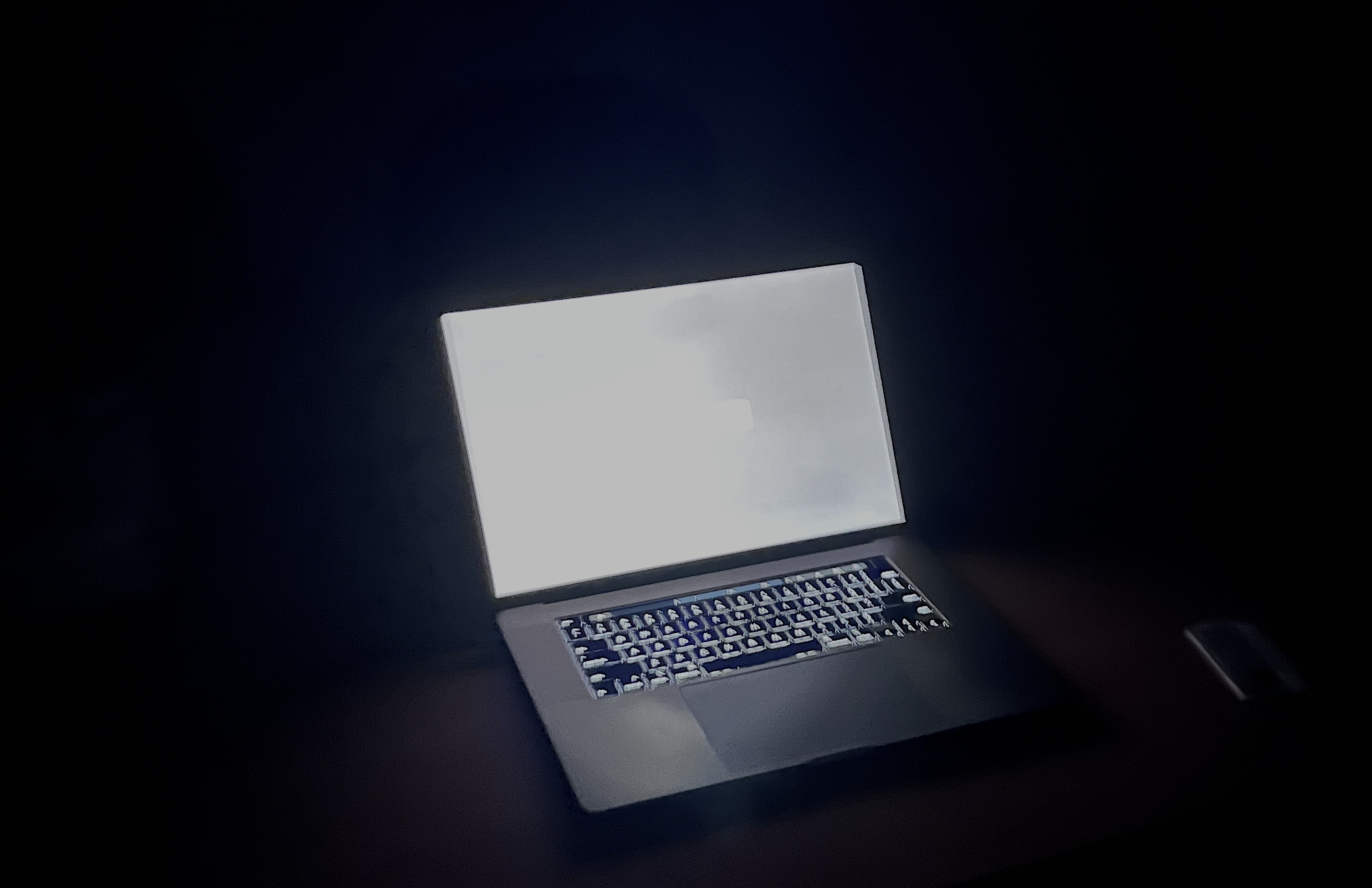}
    \subcaption{Stimuli displayed in the dark}
    \label{environmentInner}
    \end{minipage}
\caption{Laboratory Environment}
\Description{Laboratory Environment: (a) Participants were seated in front of a notebook PC. (b) Stimuli displayed in the dark.}
\end{figure}

\rev{We conduct these experiments on Day 3 (not Day 1) to ensure that participants become accustomed to the experience sampling method (ESM)-style Daily Cyberoception Measuring Survey to ensure that the Day 3 and Day 10 in-lab experiments are conducted under the same conditions.} 

On Day 10, to clarify the variation of participant's interoceptive and emotional traits during the study, we again conducted \textbf{Interoception Experiment} and \textbf{Affective Picture Rating Experiment} at the end of the experiment period.

In each in-lab experiment, the order of the experiments is randomized and well-counterbalanced. There was a break of at least 5 min (forced 5-minute rest and participant-determined further rest) between experiments to eliminate learning effects and tiredness.

\subsection{Daily Cyberoception Measuring Survey}
\label{sec:DailyCeberoceptionMeasuringSurvey}
Since the purpose of this study is to measure participants' cyberoception, the perception of various basic smartphone operations in daily use, 
an experimental environment set back to the participants' everyday lives is expected to collect objective usage data and subjective self-reports of the participants' daily use.

\subsubsection{Obtaining Participant's Subjective Perception}
\label{sec:CyberoceptionObtaining}

\begin{figure}[t]
    \centering
    \includegraphics[width=1.0\linewidth]{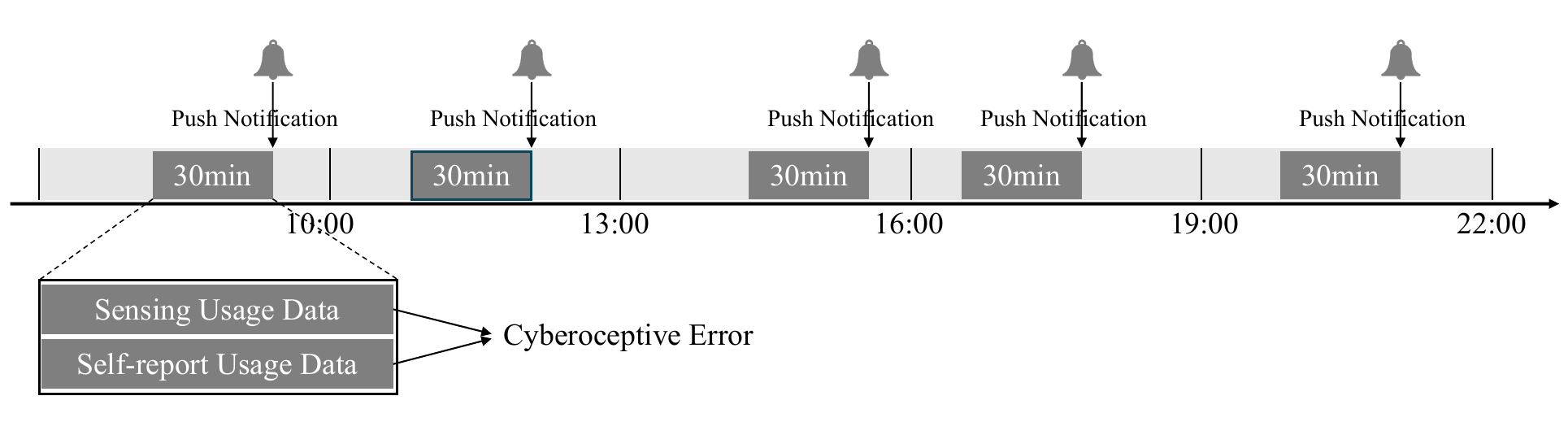}
    \caption{Daily Cyberoception Measuring Survey Procedure}
    \Description{Daily Cyberoception Measuring Survey Procedure}
    \label{fig:SurveyProcedure}

\end{figure}

In the Daily Cyberoception Measuring Survey, as shown in Figure~\ref{fig:SurveyProcedure}, we used the experience sampling method (ESM). The trends of smartphone usage vary widely throughout the day~\cite{andrews2015beyond}, and it is necessary to collect usage data from different time periods. In addition, we utilized a randomized ESM method that randomizes the timing of each survey transmission~\cite{sasaki2018comparing} in a specific time window rather than specific fixed transmission timings (noon, 2:00p.m. etc.).
We send five notifications to subjects throughout the day and collect their responses to the cyberoception metrics. Each notification is sent during five-time windows: 7:00\textasciitilde10:00, 10:00\textasciitilde13:00, 13:00\textasciitilde16:00, 16:00\textasciitilde19:00, and 19:00\textasciitilde22:00. The specific transmission time in each window is determined randomly, by delaying the actual delivery for a random period.
To measure cyberoception as a perception, we ask participants to report their subjective intuition.

Furthermore, we set the duration of each sampling session to be as short as 30 minutes immediately before answering the survey form.

To prevent participants from easily guessing the next survey questions, three of the five Cyberoception Metrics were randomly extracted for each survey form by shuffling with the Fisher-Yates algorithm \cite{durstenfeld1964algorithm}.

\subsubsection{Evaluation Metrics of Cyberoception}
\label{sec:EvaluationMetricsofCyberoception}
Our evaluation metrics of Cyberoception contain both quantitative and qualitative ones. 

\vspace{0.5cm}
\textbf{Cyberoceptive Error}: The quantitative metric of the evaluation is called \textbf{Cyberoceptive Error} shown in Equation~\ref{eq:CyberoceptionError} designed in the same way as Interoceptive Error previously presented in Section~\ref{sec:MeasurementOfInteroception}. 
Cyberoceptive Error is the calculated error rate between the self-report usage data from the participant($N_{report}$) against the sensor-data-based true usage data as ground truth($N_{real}$). 

\begin{equation}
Cyberoceptive~Error =  \frac{|N_{real}-N_{report}|}{N_{real}}
\label{eq:CyberoceptionError}
\end{equation}
\vspace{0.5cm}

\textbf{Qualitative Cyberoceptive Accuracy}: We also designed a qualitative cyberoception measure to account for the fact that participants may have difficulty quantitatively sensing smartphone use. Participants were asked to rate their smartphone usage subjectively and qualitatively by responding to a Likert scale of 1 to 5 for smartphone usage during a 30-minute session.

For instance, as the measure of participants' cyberoception on Unlocking, participants will receive a questionnaire as below:
\begin{quote}
\textit{Based on your daily experience with your smartphone use, during the period from 30 minutes ago to the present time, what do you think about your smartphone unlocking while using this smartphone? Please answer based on your own perception/feeling, and do not dwell on your answer.}
Participants can choose one item from 1 to 5, where option ``1'' represents ``No Unlocking'' and option ``5'' represents ``Most Ever''. \end{quote}
All detailed questions of the survey are listed in the Appendix~\ref{appendix:questionsInEsmSurvey}.

Participants with good cyberoception can specify the level of their smartphone usage, which means that the ascending level should correspond to usage sensing data in ascending order. Thus, to evaluate the participants' cyberoception, we calculate the Spearman correlation between the rank of Likert responses and the rank of corresponding averaged usage sensing data and named this value \textbf{Qualitative Cyberoceptive Accuracy}.

\subsubsection{System Design}
\begin{figure}[t]
    \centering
    \includegraphics[width=1.0\linewidth]{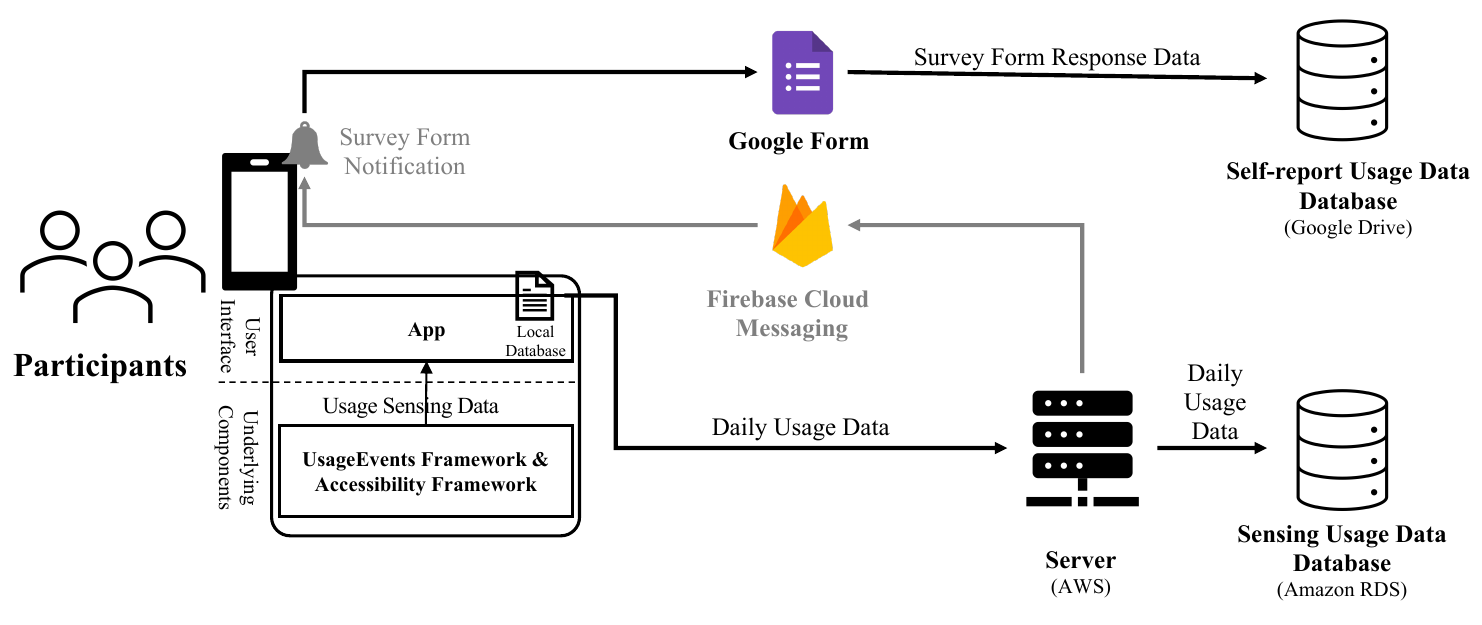}
    \caption{System Design of Daily Cyberoception Measuring Survey}
    \Description{System Design of Daily Cyberoception Measuring Survey. The experimental application has two main features, namely (1) usage data sensing and (2) self-report survey form.}
    \label{fig:SystemDesign}
\end{figure}

To collect the smartphone data and the participant's subjective answer on the context, we designed a smartphone application to be installed on each participant's smartphone.
As shown in Figure~\ref{fig:SystemDesign}, the experimental application has two main features, namely (1) usage data sensing and (2) self-report survey form.

As a concrete smartphone platform, we chose the Android platform, which allows us to easily collect various types of data from the sensors and APIs. More specifically, the application utilizes Android AppUsage Framework and Accessibility Framework to collect such data. The application is implemented as a ``foreground service'' on the Android platform to ensure that the application can collect the data continuously during the participant's smartphone daily use.
As the overall sensing framework of the smartphone application along with the data collection server (to which the data will be uploaded periodically), we used a sensing framework~AWARE \cite{ferreira2015aware, nishiyama2020ios} and its plugin~\cite{aware_appusage_plugin}.

Following the discussion in Section 5, the application collects the data regarding the six types \textbf{Cyberoception Metrics},
namely (1) Turning On, (2) Unlocking, (3) Screen Use Duration, (4) Micro-usage, (5) Most-used App, and (6) Typographical Error.
Table~\ref{tbl:cybero_val} shows the concrete data/event names used to collect the data on each metric.
To obtain the data for Metric (1) to (5), we used \texttt{UsageEvents.Event} API, and created a filter to sense five types of events: \texttt{ACTIVITY\_PAUSED}, \texttt{ACTIVITY\_RESUMED}, \texttt{SCREEN\_INTERACTIVE}, \texttt{SCREEN\_NON\_INTERACTIVE}, and \texttt{KEYGUARD\_HIDDEN}.
To collect typographical error data for Metric (6), we monitored backspace pressing activity from \texttt{Accessibility} API.

\begin{table}[t]
    \centering
    \begin{small}
    \caption{Cyberoception Metric and Corresponding Data/Event Type}
    \label{tbl:cybero_val}
    \begin{tabular}{l|c}
        \hline
        Cyberoception Metric & Sensing Event\\
        \hline\hline
        (1) Turning On & SCREEN\_INTERACTIVE\\
        \hline
        (2) Unlocking & KEYGUARD\_HIDDEN\\
        \hline
        (3) Screen Use Duration & \begin{tabular}{c}SCREEN\_INTERACTIVE\\SCREEN\_NON\_INTERACTIVE\end{tabular}\\
        \hline
        (4) Micro-usage & \begin{tabular}{c}ACTIVITY\_PAUSED\\ACTIVITY\_RESUMED\end{tabular}\\
        \hline
        (5) Most-used App & ACTIVITY\_RESUMED\\
        \hline
        (6) Typo & TYPE\_VIEW\_TEXT\_CHANGED\\
        \hline
    \end{tabular}
    \end{small}
\end{table}

\begin{itemize}

\item Turning On\\
Turning On operations are detected by the application through SCREEN\_INTERACTIVE event and stored in the usage data database.
The participants' numerical response to the Turning On metric in the survey form is collected as self-report usage data.

\item Unlocking\\
Unlocking operations are detected by the application through KEYGUARD\_HIDDEN event and stored in the database.
The participant's numerical response to the Unlocking metric in the survey form is collected as self-report usage data.

\item Screen Use Duration\\
SCREEN\_INTERACTIVE and SCREEN\_NON\_INTERACTIVE sensing events are paired into screen usage sessions. The duration of each screen usage session is calculated from the timestamps of SCREEN\_INTERACTIVE event and\\
SCREEN\_NON\_INTERACTIVE event. The accumulation of screen usage session duration within a 30-minute sampling session is treated as the sensing usage data of screen use duration.
Meanwhile, the participant's numerical response to the screen use duration metric in the survey form is treated as self-report usage data of screen use duration.

\item Micro-usage\\
ACTIVITY\_PAUSED and ACTIVITY\_RESUMED sensing events are obtained by the application as a pair. The time difference from an application's ACTIVITY\_RESUMED timestamp to the ACTIVITY\_PAUSED timestamp is calculated as a duration of one application usage. An application usage session of less than 5 s is counted and interpreted as a micro-usage.
Meanwhile, the self-report usage data of micro-usage refers to the user's numerical response to the Micro-usage metric in the survey form.

\item Most-used App\\
The information on the Most-used App for each participant was obtained during the demographic survey on Day 1.
During the data collection period, the application keeps track of and records the number of usage sessions of each application on the phone by using the outputs from the ACTIVITY\_PAUSED and ACTIVITY\_RESUMED sensing events.
Simultaneously, the participant's numerical response to the Most-used App metric in the survey form is collected as the self-report usage data.

\item Typo\\
Typo is recognized from the participant's \texttt{Delete} key-pressing activities. During the data collection period, the application continuously collects data from TYPE\_VIEW\_TEXT\_CHANGED events of Accessibility API. This event will be issued whenever the text value of a text field changes. Thus, by comparing the text value before and after such an event, the application can detect when the participant presses the Delete key.
Specifically, we calculated the Levenshtein distance and the length difference between 2 texts. When the length of text decreases, and the Levenshtein distance between 2 texts divided by the length difference is smaller than 1, the TYPE\_VIEW\_TEXT\_CHANGED will be interpreted as once Delete key pressing activity.
Meanwhile, similarly to other metrics, the participant's manual numerical response to the Typo metric in the survey form is handled as the self-report usage data on the Typo metric.
\end{itemize}

\subsection{Interoception Experiment}
\label{sec:interoceptionexperiment}
\rev{Among different modalities of interoception, we focused on the most cross-examined and extensively-researched cardiac axis.} Following the commonly-used approach in the previous literature and avoiding the floor effect in other methods ~\cite{brener1974interoceptive, hart2013emotional}, we measured participants' Interoceptive Error (presented in Section~\ref{sec:MeasurementOfInteroception}) by the heartbeat counting task (HCT) that evaluates the participants' cardiac perception ability to their heartbeats. 
We calculate Interoceptive Error based on the heart rate figure from the ECG sensor (as the ground truth) ($N_{real}$), and the self-report number from the participant($N_{report}$). For our experiment, used Polar H10 N ECG sensor~\cite{polarH10N}.  



\begin{figure}[t]
    \centering
    \includegraphics[width=\linewidth]{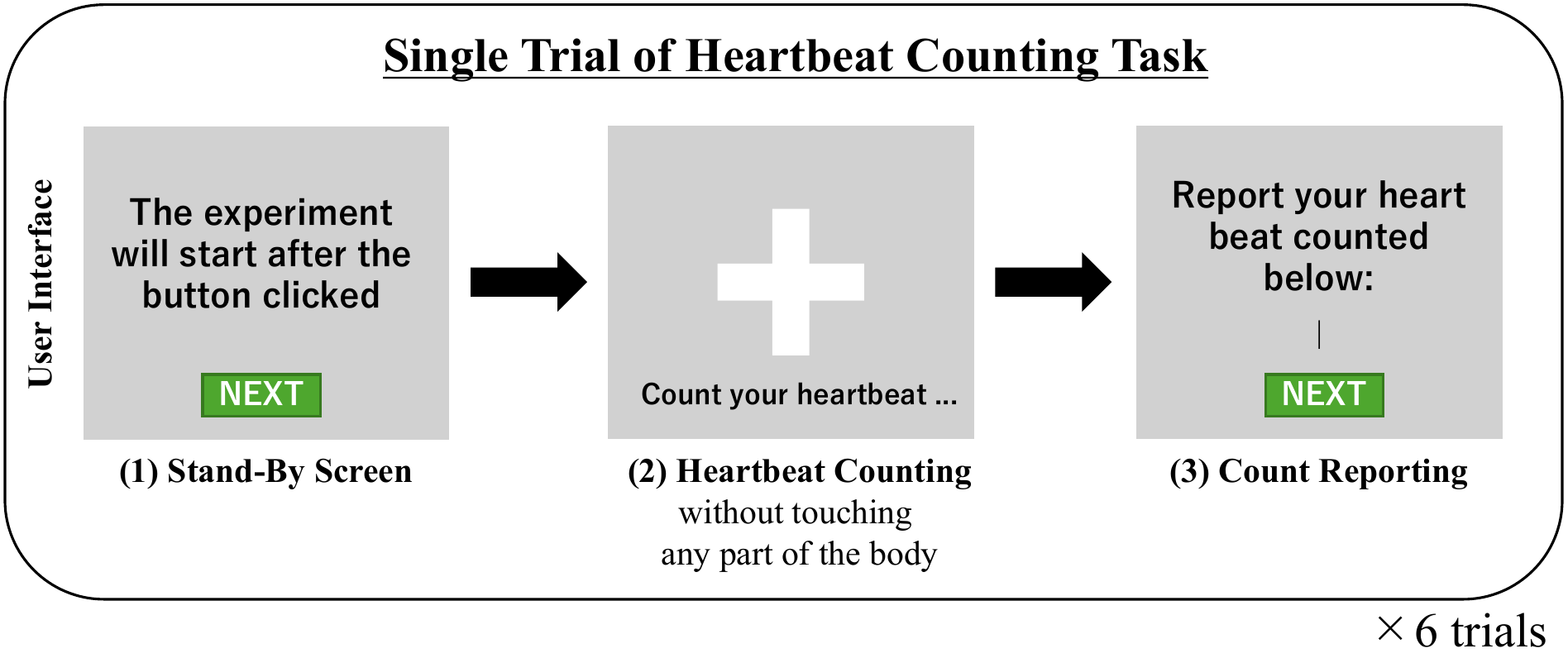}
    \caption{Procedure of Heartbeat Counting Task}
    \Description{Procedure of Heartbeat Counting Task. 1 trial consists of 3 steps, literally (1) Stand-By, (2) Heartbeat Counting, and (3) Count Reporting. Following the user interface on the PC screen, each participant proceeds the steps from (1) to (3).  
    Each participant will do six trials.}    
    \label{fig:HCT}
\end{figure}

\subsubsection{Procedure}
Figure~\ref{fig:HCT} shows the overview of the heartbeat counting task. 1 trial consists of 3 steps, literally (1) Stand-By, (2) Heartbeat Counting, and (3) Count Reporting. Following the user interface on the PC screen, each participant proceeds the steps from (1) to (3). 
Each participant will do six trials. 

The duration of (2) heartbeat counting steps in the six trials are 25 seconds, 30 s, 35 s, 40 s, 45 s, and 50 s, respectively. The information on these time durations is not told to the subjects, making it difficult for them to guess the heart rate from the time duration. (For example, when the participant is asked to count the heartbeat for 60 s, they can actually estimate the number based on the knowledge and common sense about the human's typical heart rate.)
The order of the trials is rearranged randomly to ensure that subjects cannot easily guess the time duration of each trial.
Subjects are asked to count their own heartbeat without touching their wrists or any other body part.
They are also instructed to keep their hands on the desk during the experiment.

\subsection{Affective Picture Rating Experiment}
\label{sec:affectivepicture}
The participant's personal affective traits are measured in the Affective Picture Rating Experiment.
The Affective Picture Rating Experiment presents emotional images to the participant in the experiment.
Subsequently, the participant rates them on two dimensions of emotional experience: emotional valence and arousal levels.
The same stimulus, an image, is used to evoke the same emotional state in the participants.

\subsubsection{Stimulus}
The experiment in this study is designed with reference to the experiment by Lang, P.J., et al~\cite{lang2005international}, and the International Affective Picture System (IAPS) is used as the stimulus. 

\rev{The IAPS is useful for assessing state emotion but can also be applied to evaluate trait emotion. Many prior studies, particularly those exploring the relationship between trait emotion and interoception, have employed the IAPS~\cite{parrinello2022embodied}. In this study, we analyze trait emotion by exposing different participants to the same controlled environment and evaluating their emotional responses. Here, the IAPS is valuable for maintaining a consistent environment.}

This study created a subset from IAPS in which pictures rated inconsistently are excluded (standard deviation larger than 2). Furthermore, images with a high possibility of violating ethics were subjectively excluded. After the exclusion, stimuli were extracted in consideration of balance: 8 images for each of 9 categories of valence/arousal values, (1) low-valence low-arousal, (2) low-valence neutral-arousal, (3) low-valence high-arousal, (4) neutral-valence low-arousal, (5) neutral-valence neutral-arousal, (6) neural-valence high-arousal, (7) high-valence low-arousal, (8) high-valence neutral-arousal, and (9) high-valence high-arousal.
\rev{The valence labeling score of the subset ranges from 2.43 to 7.57(M = 4.98, SD = 1.33). The arousal labeling score of the subset ranges from 1.72 to 6.99(M = 4.48, SD = 1.29).}

\subsubsection{Scale}
\begin{figure}[t]
    \begin{minipage}[b]{1.0\linewidth}
    \centering
    \includegraphics[width=\linewidth]{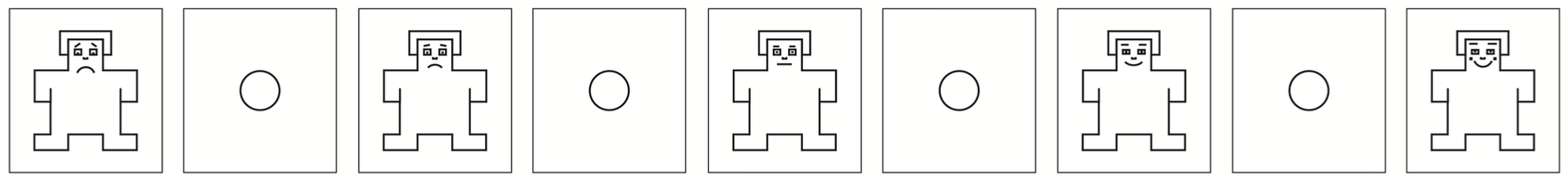}
    \subcaption{\begin{tabular}{c}Valence Dimension\end{tabular}}
    \label{fig:SAM_valence}
    \end{minipage}\\
    \vspace{0.5cm}
    \begin{minipage}[b]{1.0\linewidth}
    \centering
    \includegraphics[width=\linewidth]{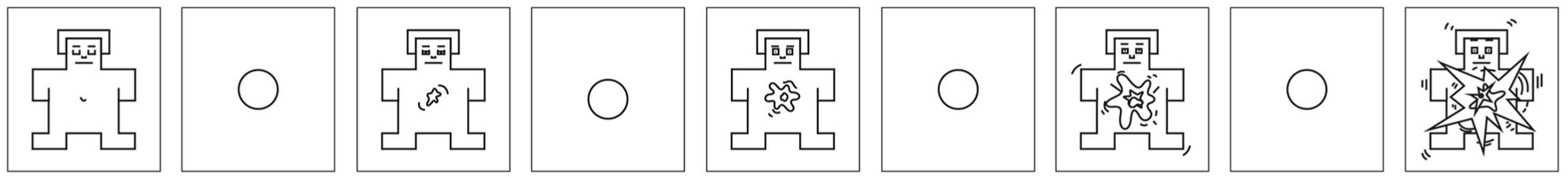}
    \subcaption{\begin{tabular}{c}Arousal Dimension\end{tabular}}
    \label{fig:SAM_arousal}
    \end{minipage}\\
\caption{S.A.M Scale Used to Evaluate Paticipant's Emotional Experiences}
\Description{S.A.M Scale Used to Evaluate Paticipant's Emotional Experiences}
\label{fig:SAM}
\end{figure}

Self assessment Manikin (S.A.M) (Fig. \ref{fig:SAM}) is used in the affective picture rating task.
S.A.M. is a scale for measuring emotion proposed by Lang~\cite{bradley1994measuring} and is widely used in emotion estimation research. \rev{The reliability and validity of S.A.M. have been demonstrated in numerous studies. For instance, both within- and
between-subject reliability is reported by Lang~\cite{lang1997international}. The validity of S.A.M, particularly in research on interception and emotion, is also supported by numerous studies~\cite{parrinello2022embodied}. 
The present study uses the two dimensions of the S.A.M., emotional valence and arousal, corresponding to the core affect model.}
In the S.A.M, the five human figures and the four circles in between are arranged in two sets vertically and continuously.
The emotional valence dimension scale represents positive and negative emotions (also known as pleasant and unpleasant).
As shown in Fig. \ref{fig:SAM_valence}, the scale varies from a smile to a frown. 
The scale of the arousal dimension represents the emotions of excitement and calmness.
As shown in Fig. \ref{fig:SAM_arousal}, the left pole represents relaxed, non-aroused emotion. The right pole represents excited and aroused emotions.

The subjects are asked to choose the option that best matches their current emotional experience. The vertical order of the two scales is randomized.
The selected option is converted into numerical data from 1 to 9.

\subsubsection{Procedure}
\begin{figure}[t]
    \centering
    \includegraphics[width=\linewidth]{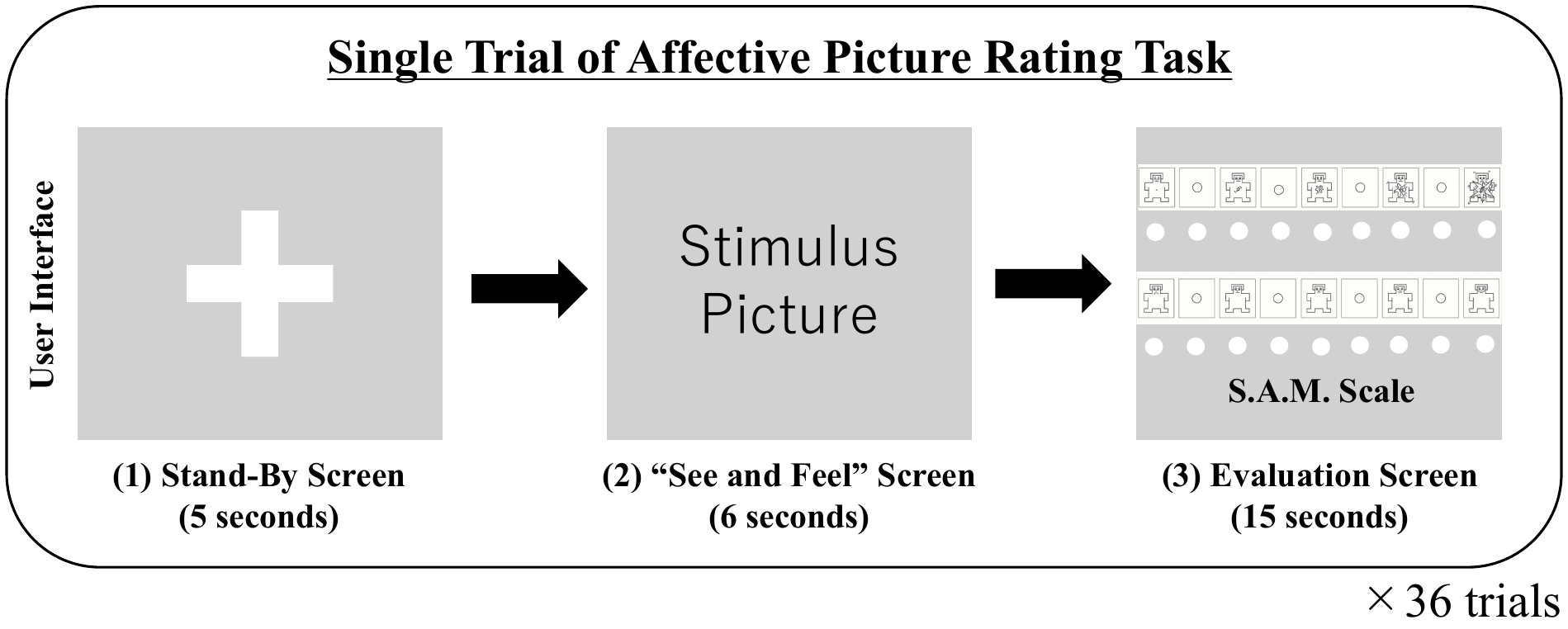}
    \caption{Procedure of Affective Picture Rating Task}
    \Description{Procedure of Affective Picture Rating Task. Each trial of the Affective Picture Rating Task consists of screens of three steps: (1) Stand by, (2) ``See and Feel'', and (3) Evaluation.}
    \label{fig:EmotionalPictureRatingTask}
\end{figure}

As shown in Fig.~\ref{fig:EmotionalPictureRatingTask}, each trial of the Affective Picture Rating Task consists of screens of three steps: (1) Stand by, (2) ``See and Feel'', and (3) Evaluation.
First, a 5-second preparation period is given before each picture is presented. During this time, a white cross is presented. In the (2) ``See and Feel'' period, each affective picture is presented for 6 s. After that, the screen switches to a 15-second (3) ``Evaluation'' screen. The evaluation screen uses a two-dimensional S.A.M. scale with nine choices for each dimension. The subjects are instructed to observe the facial expressions of the scale mannequins before answering the questions. 
These trials are repeated 36 times for each participant.

\subsection{Typing Experiment}
\label{sec:typingexperiment}
To acquire the participant's sense of Typographical Error in an in-lab environment, we conduct \textbf{Typing Experiment}.

\subsubsection{Content}
We selected text content to be typed from general news articles.
The criterion for article selection is that the content is as ordinary as possible to minimize the variance in the difficulty of understanding among the participants. In this study, we chose a short text on the topic of weather, with a length of 253 words (1376 keystrokes).

\subsubsection{Procedure}
We show the content on a piece of paper and ask each participant to type on a mailing application on the smartphone.
During the experiment, auto-capitalization, auto-correction, and predictive-text features of the keyboard on the phone are turned off (the instruction to do so is given by the experimenter to the participant) and are also prohibited from being used.
In other words, participants are set in an environment where they must press every keystroke to input the displayed text.

\subsection{Novelty of the Methodology}
\rev{The proposed methodology incorporates a novel integration of interoceptive and cyberoceptive measures to understand emotional traits. Unlike traditional methods that rely solely on self-reported data or physiological measures, our approach combines subjective perception metrics (cyberoception) with objective in-lab experiments designed to measure emotional traits and interoceptive abilities. This dual approach enables a more comprehensive evaluation of the relationship between daily smartphone usage and emotional states. }

\rev{The novelty of our method also lies in the randomized ESM combined with a smartphone application that continuously collects real-world interaction data. By employing statistic analysis methods, we quantitatively measure the cyberoception of participants. Additionally, the in-lab experiments are carefully designed to assess emotional and interoceptive traits, allowing for a robust analysis of their relationship with daily cyberoception.}

\section{Participants}
The participants were recruited inside our university, where the experiment was approved by the IRB (Institutional Review Board).
The recruitment was done via email from the university office to all students in our information science department.
Twenty-five participants majoring in information science and related disciplines were recruited.

All participants are daily smartphone (Android OS version 9 or above) users.
The participants were paid SGD60 for their full participation.

At the study briefing, the participants answered demographic survey forms after obtaining informed consent.

\subsection{Demographics}
Twenty-five participants (19 men and 6 women) range in age from 19 to 34 years (M = 23.1, SD = 2.9). 
\rev{While the number of participants is modest, participants were recruited from a homogeneous group with similar age ranges and backgrounds, which minimizes variability in the dataset. Given the low inter-participant variability, it is feasible to conduct a statistically valid analysis with a relatively small sample size.}

Most participants were using English while typing on their smartphones (Always = 20, Usually = 3, Often = 1, Sometimes = 1).


Auto-correction refers to the feature implemented in the Android keyboard, which can correct typos, including capitalization errors, misspelled words, and missing pieces of text.
Fifteen participants were using the auto-correction feature in high frequency (Always, Usually, and Often) while typing on their smartphones. 
The predictive text refers to another feature in the keyboard, which predicts and suggests the text the user may wish to insert. 
19 participants were using the predictive text features at low frequency (Sometimes, Hardly ever, and Never).

Brands of participants' smartphones include Samsung (N = 12), OPPO (N = 3), Google (N = 3), One Plus (N = 2), Xiaomi (N = 2), Sony (N = 1), Poco (N = 1), and Huawei (N = 1). Android versions of the smartphones include 13 (N = 19), 12 (N = 3), 11 (N = 1), 10 (N = 1), and EMUI 12 (Android-based mobile operating system of Huawei smartphone, N = 1).

Individually, the most frequently used applications are collected from the participants' subjective reports: Telegram (N = 12), TikTok (N = 3), Instagram (N = 2), Whatsapp (N = 2), Twitter (N = 1), Pinterest (N = 1), YouTube (N = 1), Netflix (N = 1), WeChat (N = 1), Google News (N = 1).

\section{Acquired Data}
Out of 25 participants, one received an invalid survey form due to a network communication problem, and two reported their screen usage duration longer than 30 minutes within 30 minutes, whose data are labeled as low credibility. Data from the above participants are excluded from data analysis, and the remaining data from the other 22 are used.

The acquired data include data from 2 parts: Daily Cyberoception Measuring Survey and In-lab Experiments (Interoception Experiment, Affective Picture Rating Experiment, and Typing Experiment).

\subsection{Daily Cyberoception Measuring Survey}
Pairs of subjective self-report responses and corresponding sensor data from smartphone usage constitute the data collected from the Daily Cyberoception Measuring Survey. We collected 259 data points of Turning On (Mean = 11.78, SD = 6.03), 286 data points of Unlocking (Mean = 13, SD = 7.59 ), 258 data points of screen use duration (Mean = 12.90, SD = 5.32), 305 data points of micro-usage (Mean = 6.99, SD = 9.37), 219 data points of Typo (Mean = 9.95, SD = 4.78) and 98 data points of Most-used App (Mean = 4.45, SD = 5.09).
In total, 1425 valid self-report responses and 10-day smartphone usage data are collected from 22 participants.

\subsection{In-Lab Experiments}
During the Interoception Experiment, participants' average heart rates ranged from 58.65 to 98.64 (M = 74.52, SD = 10.95).

In the Affective Picture Rating Experiment mentioned, the participants rated affective pictures from two aspects: arousal and valence. 
Their average valence rating scores range from 3.56 to 5.80 (M = 5.12, SD = 0.51), and the average arousal rating scores range from 1.00 to 6.28 (M = 3.79, SD = 1.34).

In the Typing Experiment, participants' delete key pressing count ranges from 0 to 304 (M = 102.8, SD = 77.95).
\section{Results}

\subsection{Sensing Usage Data}
On the six types of Sensing Usage Data collected from each participant's smartphone API, as ``objective'' observation on each Cyberoception Metric, here we introduce some statistical summary, such as distribution and variation among the participants.

\subsubsection{Distribution}
\begin{figure}[t]
\vspace{-0.2cm}
    \begin{minipage}[b]{0.48\linewidth}
    \centering
    \includegraphics[width=1.1\linewidth]{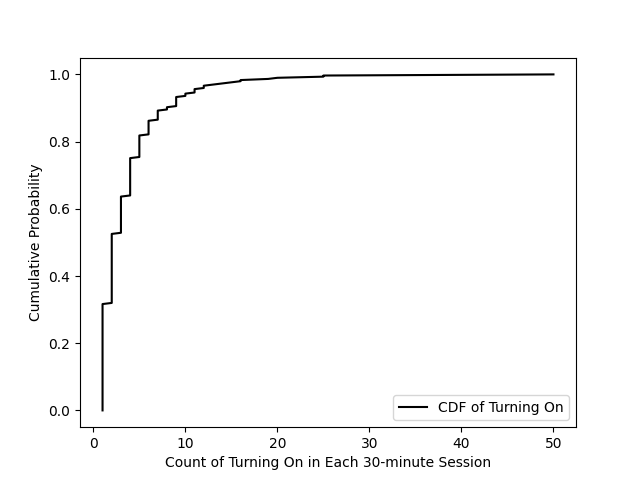}
    \subcaption{\begin{tabular}{c}Turning On\end{tabular}}
    \end{minipage}
    \begin{minipage}[b]{0.48\linewidth}
    \centering
    \includegraphics[width=1.1\linewidth]{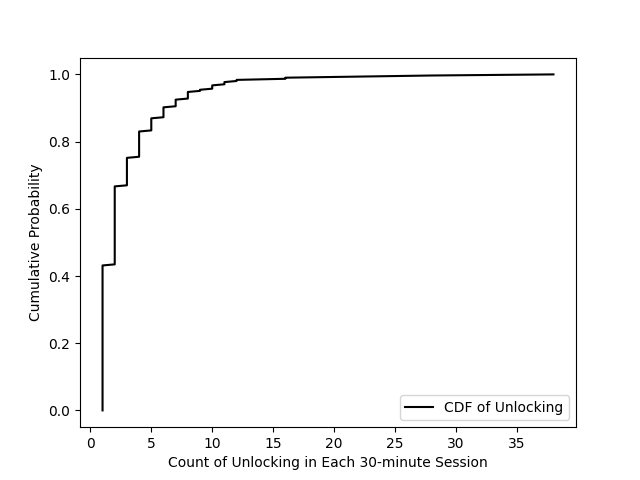}
    \subcaption{\begin{tabular}{c}Unlocking\end{tabular}}
    \end{minipage}\\
    \begin{minipage}[b]{0.48\linewidth}
    \centering
    \includegraphics[width=1.1\linewidth]{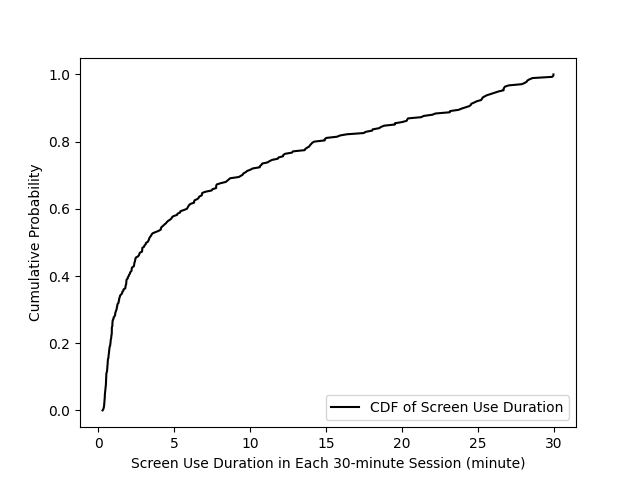}
    \subcaption{\begin{tabular}{c}Screen Use Duration\end{tabular}}
    \end{minipage}
    \begin{minipage}[b]{0.48\linewidth}
    \centering
    \includegraphics[width=1.1\linewidth]{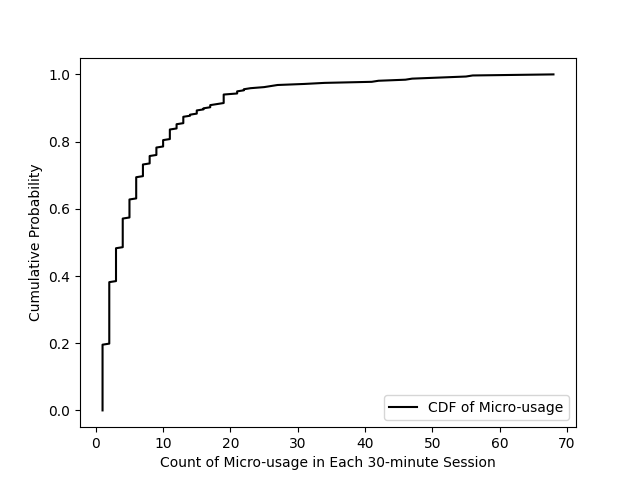}
    \subcaption{\begin{tabular}{c}Micro-usage\end{tabular}}
    \end{minipage}\\
    \begin{minipage}[b]{0.48\linewidth}
    \centering
    \includegraphics[width=1.1\linewidth]{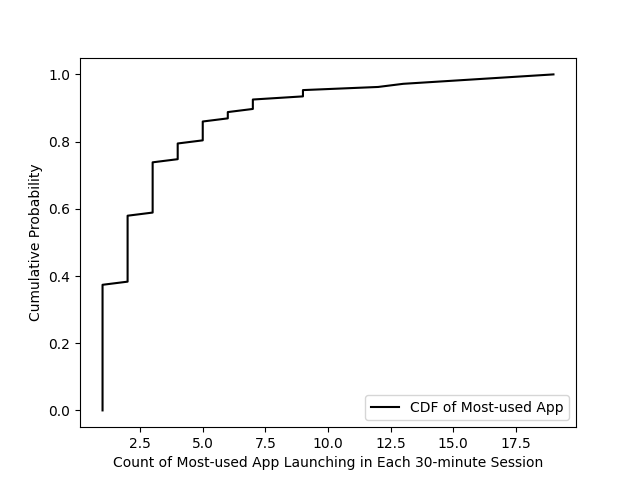}
    \subcaption{\begin{tabular}{c}Most-used App \end{tabular}}
    \end{minipage}
    \begin{minipage}[b]{0.48\linewidth}
    \centering
    \includegraphics[width=1.1\linewidth]{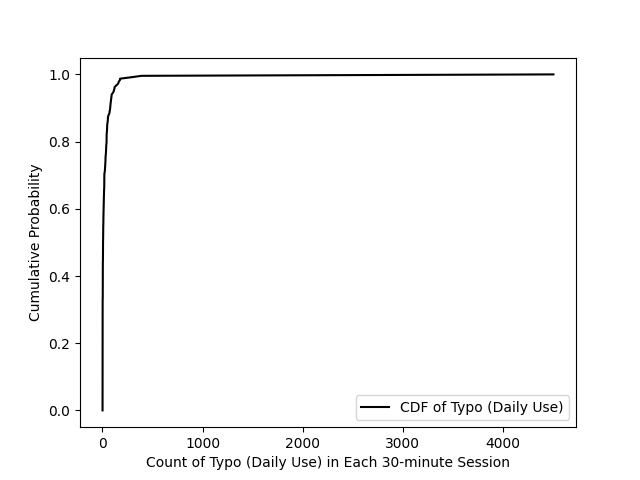}
    \subcaption{\begin{tabular}{c}Typo(Daily Use)\end{tabular}}
    \end{minipage}\\
\vspace{-0.3cm}
\caption{Distribution of Usage Sensing Data}
\Description{Distribution of Usage Sensing Data}
\label{fig:RawDataDistribution}
\vspace{-0.4cm}
\end{figure}

The distribution of each type of Cyberoception Metric sensing data during the 30-minute sessions is shown in Figure~\ref{fig:RawDataDistribution}. 

\subsubsection{Variation among Subjects}


The range of the collected value, along with the mean and standard deviation among the participants are as follows.
Turning On ranges from 1.57 to 8 (Mean = 3.83, SD = 1.55, N = 19); 
Unlocking ranges from 1.50 to 5.07 (Mean = 3.03, SD = 1.18, N = 19); 
Screen Use Duration ranges from 0.94 to 18.60 (Mean = 8.08, SD = 4.11, N = 20); 
Micro-usage ranges from 2.53 to 51.67 (Mean = 8.82, SD = 10.02, N = 20); 
Most-used App ranges from 1.67 to 6.31 (Mean = 2.77, SD = 1.31, N = 11); 
and Typo ranges from 4.85 to 52.24 (Mean = 23.68, SD = 15.90, N = 21).

\subsubsection{Correlation between the Operations}
We conducted a correlation analysis between the different operations. 
Between the frequencies of Turning On and Unlocking, we saw a high correlation (r = 0.70, p < 0.001), showing that Turning On operations are very often accompanied by following the Unlocking operation.



\subsection{Cyberoception}

\aptLtoX[graphic=no,type=html]{\begin{table*}
    \centering
    \caption{Correlation between Participants' Cyberoceptive Error (CE) and Qualitative Cyberoceptive Accuracy (QCA)}
    \vspace{-0.2cm}
    \label{tbl:CE_QCA}
      \begin{tabular}{c|cccccc}
        \hline
        & Turning On & Unlocking & Screen Use Duration & Micro-usage & Most-used App & \begin{tabular}{c}Typo\\(daily-life)\end{tabular}\\
        \hline\hline
        \begin{tabular}{c}$r$-value\\$p$-value\end{tabular}& \begin{tabular}{c}-0.373\\(0.116)\end{tabular} & \begin{tabular}{c}-0.029\\(0.907)\end{tabular} & \begin{tabular}{c}-0.384\\(0.104)\end{tabular} & \begin{tabular}{c}-0.012\\(0.958)\end{tabular} & \begin{tabular}{c}-0.606\\(0.048)\end{tabular} &\begin{tabular}{c}0.691\\(0.001)\end{tabular}\\
        \hline
    \end{tabular}
\end{table*}}{\begin{table*}[t]
    \centering
    \caption{Correlation between Participants' Cyberoceptive Error (CE) and Qualitative Cyberoceptive Accuracy (QCA)}
    \vspace{-0.2cm}
    \label{tbl:CE_QCA}
    \scalebox{0.85}[0.9]{
    \begin{tabular}{c|cccccc}
        \hline
        & Turning On & Unlocking & Screen Use Duration & Micro-usage & Most-used App & \begin{tabular}{c}Typo\\(daily-life)\end{tabular}\\
        \hline\hline
        \begin{tabular}{c}$r$-value\\$p$-value\end{tabular}& \begin{tabular}{c}-0.373\\(0.116)\end{tabular} & \begin{tabular}{c}-0.029\\(0.907)\end{tabular} & \begin{tabular}{c}-0.384\\(0.104)\end{tabular} & \begin{tabular}{c}-0.012\\(0.958)\end{tabular} & \begin{tabular}{c}-0.606\\(0.048)\end{tabular} &\begin{tabular}{c}0.691\\(0.001)\end{tabular}\\
        \hline
    \end{tabular}}
\end{table*}}

\subsubsection{Quantitative Cyberoceptive Error}
Cyberoceptive Error represents how accurately (or inaccurately) each participant perceives their smartphone usage, to evaluate participants' Cyberoceptive ability (Equation \ref{eq:CyberoceptionError}). 

As mentioned in ~\ref{sec:CyberoceptionObtaining}, the participants were required to answer survey forms sent by notifications several times a day. 
The questions in each form are randomly selected from six types of cyberoception metrics. 
For some metrics, only a few valid data were successfully collected occasionally since the participants might have missed some responses to the notifications. 
For each participant, after excluding metrics that were responded less than three times, the Cyberoceptive Errors of six metrics were calculated. 

\subsubsection*{Variation among Subjects}
The range of the calculated Cyberoceptive Errors (for each participant), along with the mean and standard deviation among the participants, are as follows.

Cyberoceptive Error of Turning On ranges from 0.51 to 3.65 (M = 1.49, SD = 0.93, N = 19); 
Cyberoceptive Error of Unlocking ranges from 0.49 to 2.03 (M = 1.08, SD = 0.42, N = 19); 
Cyberoceptive Error of Screen Use Duration ranges from 0.40 to 3.94 (M = 1.45, SD = 0.96, N = 20); 
Cyberoceptive Error of Micro-usage ranges from 0.52 to 1.31 (M = 0.90, SD = 0.17, N = 20); 
Cyberoceptive Error of Most-used App ranges from 0.62 to 1.38 (M = 0.81, SD = 0.26, N = 11); 
Cyberoceptive Error of Typo(daily-life) ranges from 0.61 to 1.01 (M = 0.63, SD = 0.21, N = 21); 
Cyberoceptive Error of Typo(in-lab) ranges from 0 to 304 (M = 102.82, SD = 77.95, N = 22).

Among the participants, the cyberoception values vary. 
Distribution of Turning On Cyberoceptive Error, Unlocking Cyberoceptive Error, Most-used App Cyberoceptive Error, and Screen Use Duration Cyberoceptive Error are negatively skewed. Meanwhile, the distribution of Micro-usage Cyberoceptive Error and Typo (in-lab) Cyberoceptive Error are slightly positively skewed. Typo (daily-life) Cyberoceptive Error seems to follow a normal distribution, while Typo (in-lab) Cyberoceptive Error is lower in general. 

\subsubsection*{Correlation between the Metrics}
Correlational analysis among Cyberoceptive Errors of six metrics was performed. 
Although Turning On is often accompanied by an Unlocking operation, the Cyperoceptive Error of Unlocking and Cyperoceptive Error of Turning On are found to be hardly correlated (r = 0.22, p = 0.36). 
Overall correlations among Cyberoceptive Errors of metrics are not significant but positive. 

\begin{figure}[t]
    \centering
    \includegraphics[width=\linewidth]{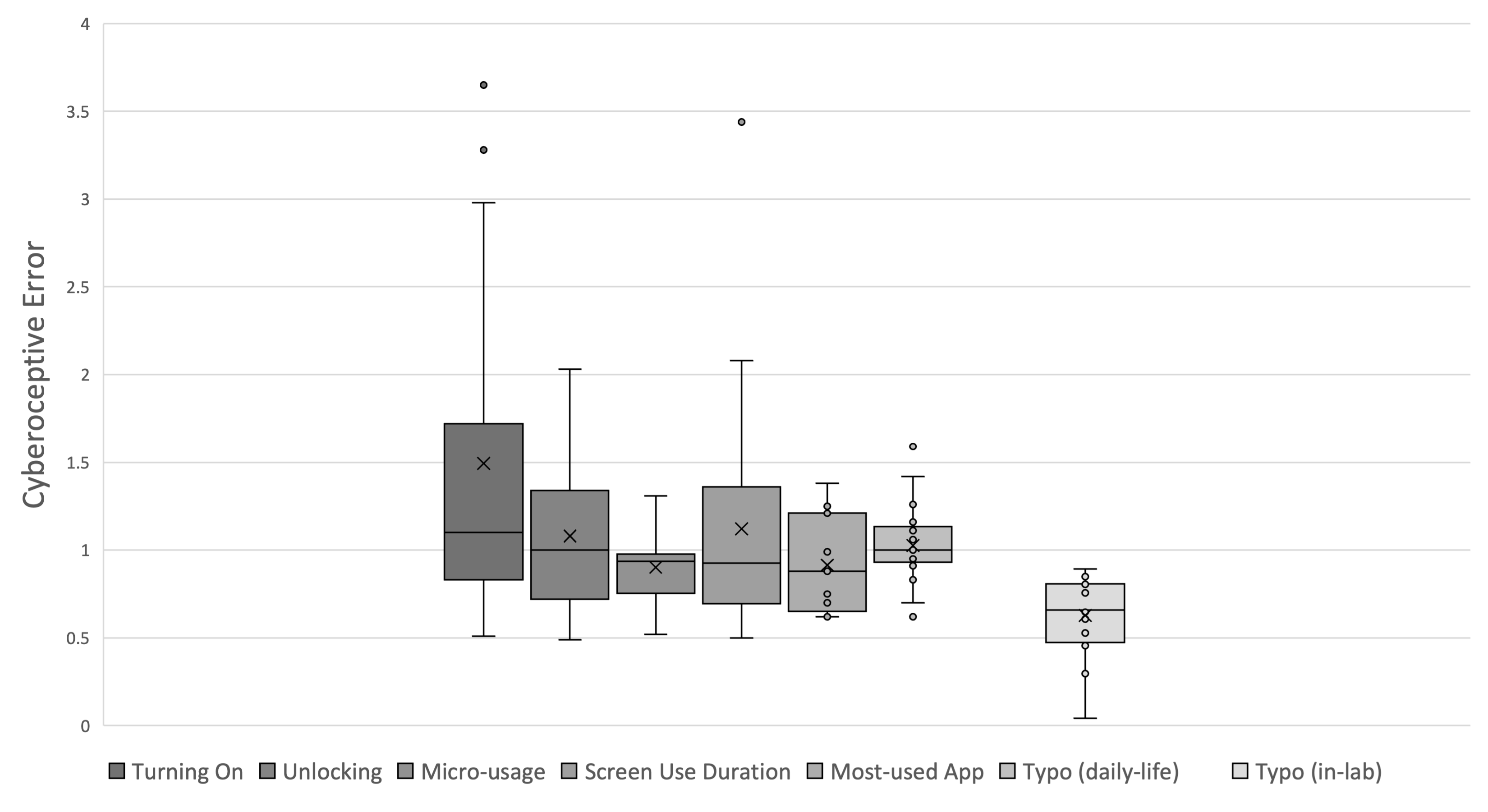}
    \vspace{-0.7cm}
    \caption{Box Plot of Each Cyberoception Metric}
    \Description{Box Plot of Each Cyberoception Metric}
    \label{fig:CyberoceptionItemsBoxplot}
    \vspace{-0.3cm}
\end{figure}

\begin{figure}[t]
    \begin{minipage}[b]{0.49\linewidth}
    \centering
    \includegraphics[width=\linewidth]{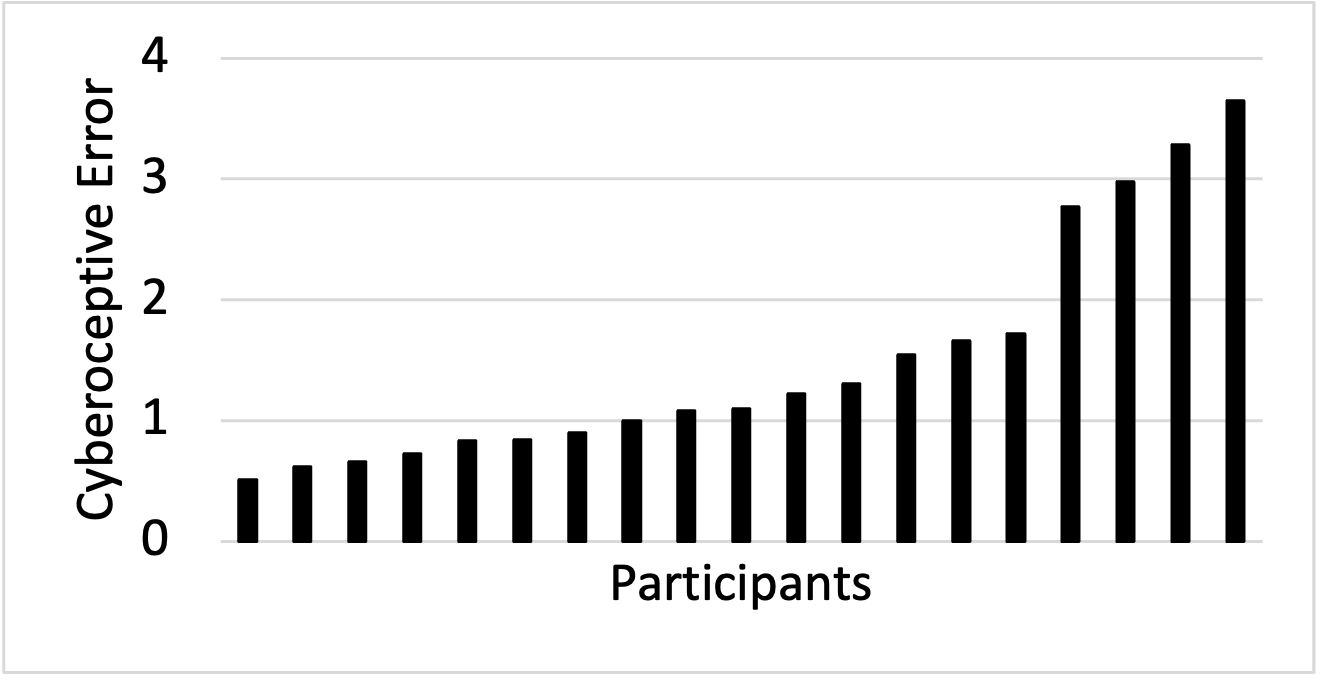}
    \subcaption{\begin{tabular}{c}Turning On\end{tabular}}
    \label{fig:TurningOn}
    \end{minipage}
    \begin{minipage}[b]{0.49\linewidth}
    \centering
    \includegraphics[width=\linewidth]{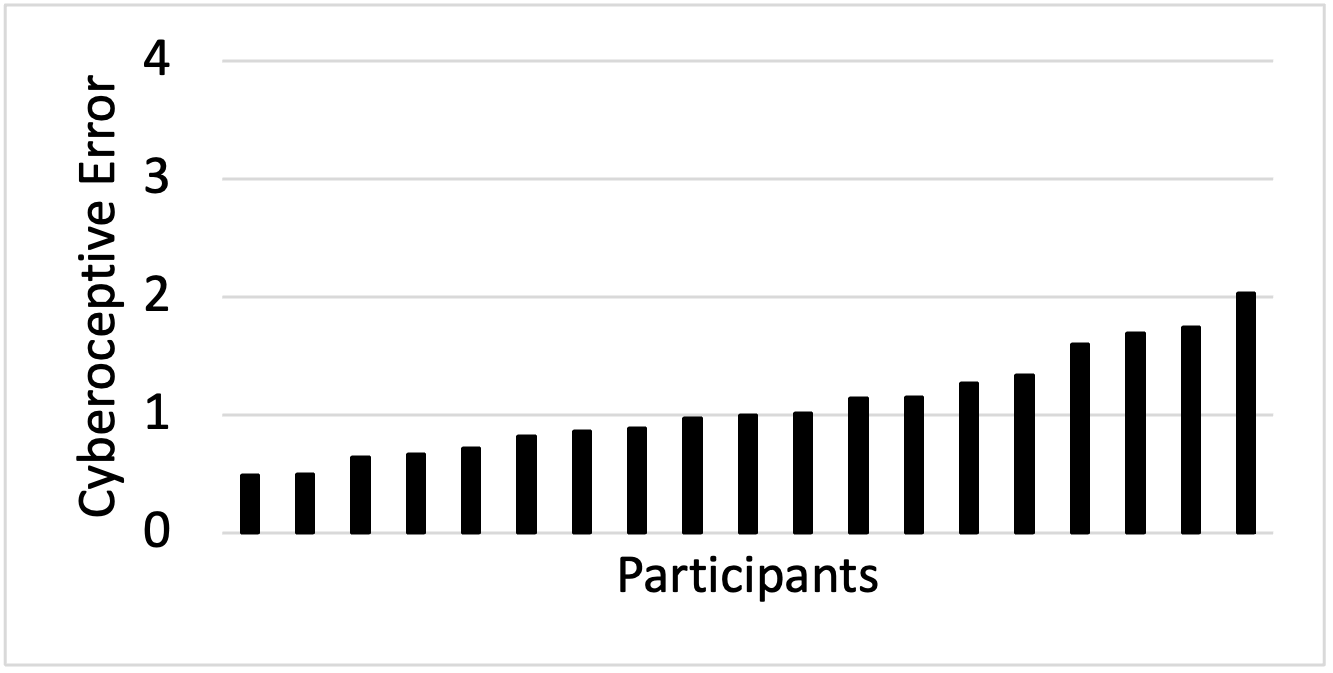}
    \subcaption{\begin{tabular}{c}Unlocking\end{tabular}}
    \label{fig:Unlocking}
    \end{minipage}\\

    \begin{minipage}[b]{0.49\linewidth}
    \centering
    \includegraphics[width=\linewidth]{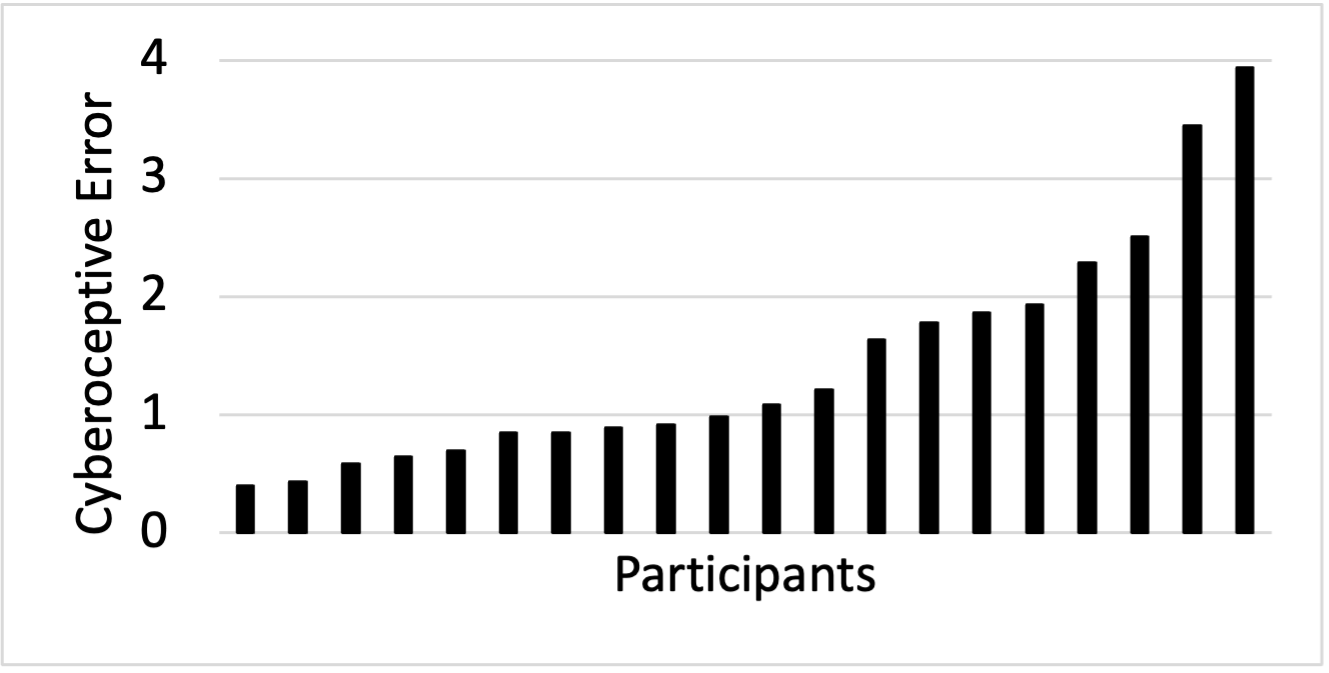}
    \subcaption{\begin{tabular}{c}Screen Use Duration\end{tabular}}
    \label{fig:ScreenUseDuration}
    \end{minipage}
    \begin{minipage}[b]{0.49\linewidth}
    \centering
    \includegraphics[width=\linewidth]{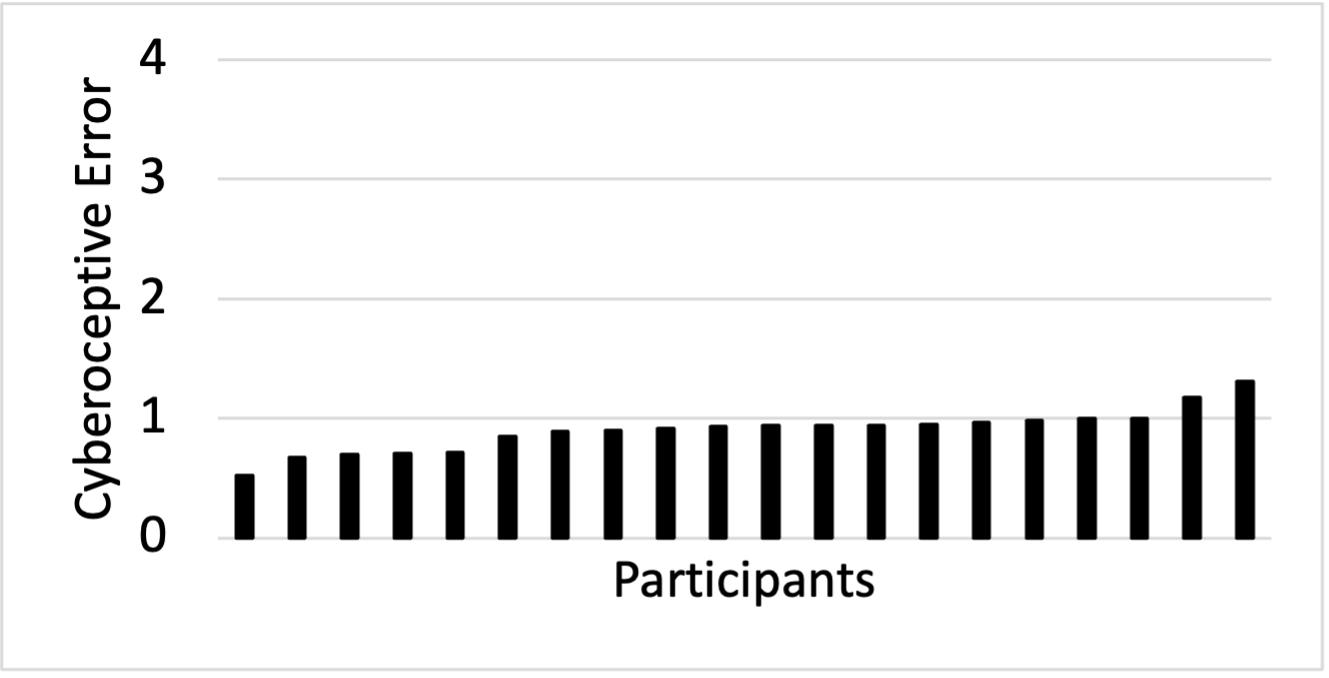}
    \subcaption{\begin{tabular}{c}Micro-usage\end{tabular}}
    \label{fig:MicroUsage}
    \end{minipage}\\

    \begin{minipage}[b]{0.49\linewidth}
    \centering
    \includegraphics[width=\linewidth]{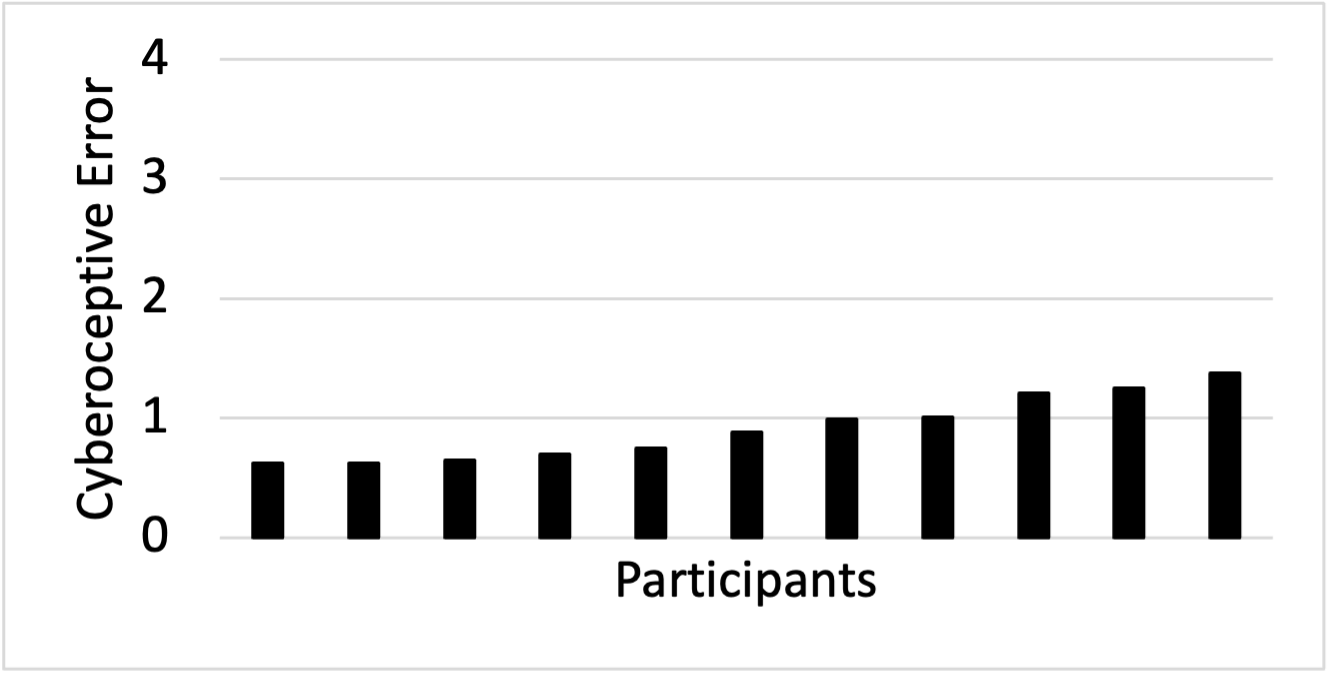}
    \subcaption{\begin{tabular}{c}Most-used App\end{tabular}}
    \label{fig:MostUsedApp}
    \end{minipage}
    \begin{minipage}[b]{0.49\linewidth}
    \centering
    \includegraphics[width=\linewidth]{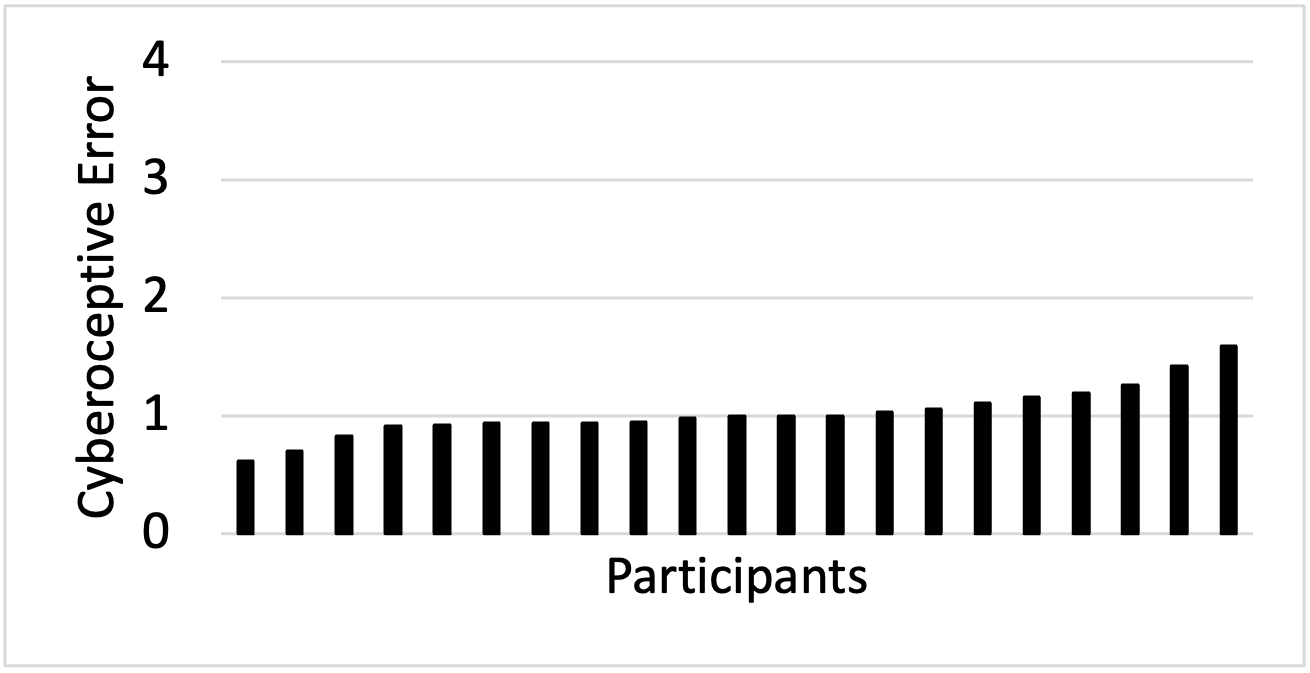}
    \subcaption{\begin{tabular}{c}Typo (daily-life)\end{tabular}}
    \label{fig:TypoDailyLife}
    \end{minipage}\\

    \begin{minipage}[b]{0.49\linewidth}
    \centering
    \includegraphics[width=\linewidth]{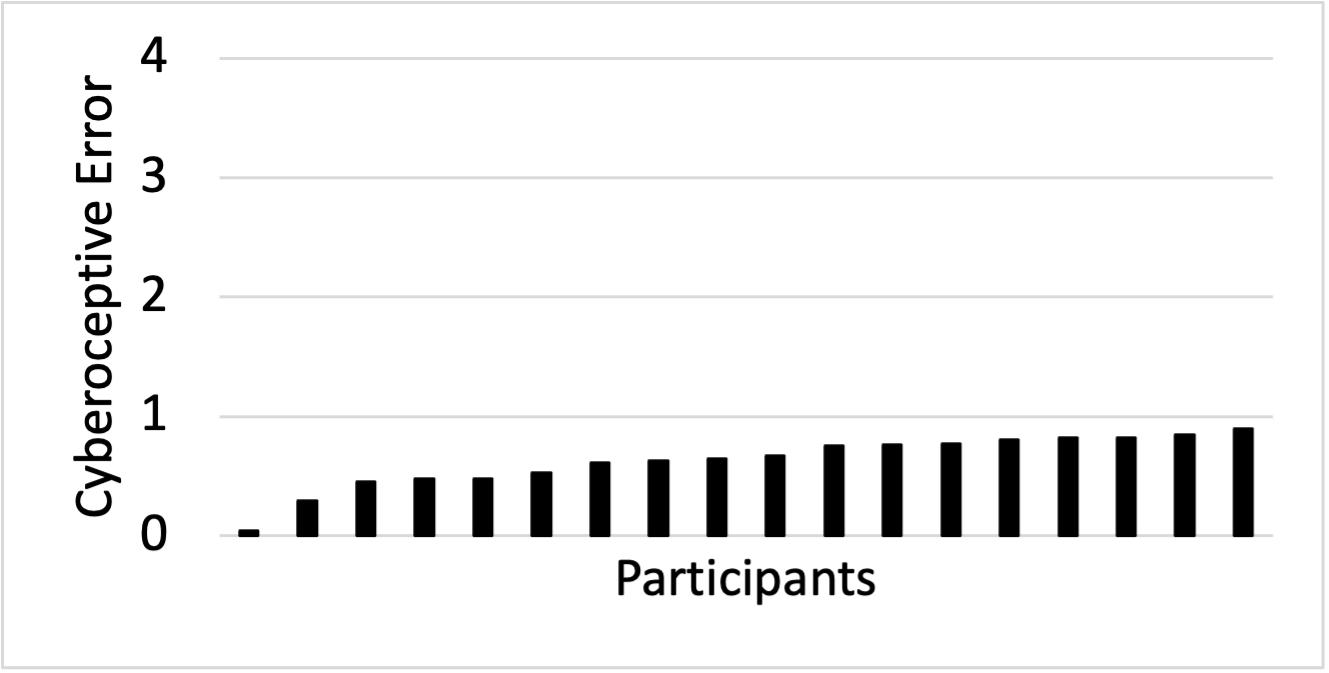}
    \subcaption{\begin{tabular}{c}Typo (in-lab)\end{tabular}}
    \label{fig:TypoInLab}
    \end{minipage}\\
\vspace{-0.2cm}
\caption{Individual Difference of Participants' Cyberoceptive Error on Cyberoception Metrics}
\Description{Individual Difference of Participants' Cyberoceptive Error on Cyberoception Metrics}
\vspace{-0.3cm}
\end{figure}

\subsubsection{Qualitative Cyberoceptive Accuracy}
Qualitative Cyberoceptive Accuracy represents participants' Cyberoceptive ability, calculated from the collected sensor data and the participant's subjective report on a Likert-type scale. 
Participants with low (quantitative) Cyberoceptive Error are expected to have high qualitative cyberoceptive accuracy. 
To evaluate the degree of concordance between the two measurements, we conducted a Pearson correlation between (quantitative) Cyberoceptive Error (CE) and Qualitative Cyberoceptive Accuracy (QCA), as shown in Table~\ref{tbl:CE_QCA}.

\subsection{Interoception}
Based on the results of the participants' interoception collected through the in-lab experiment, most participants performed highly accurate interoception (M = 0.32, SD = 0.11). 21 out of 22 participants counted their heartbeat correctly with an Interoceptive Error lower than 0.5. 


\subsection{Gender Difference}
Despite the unbalanced amount of collected data between men and women, we attempted to analyze differences in the results between genders using ANOVA. The result shows that men tended to report higher emotional arousal than women ($M_{women}=0.394$, $M_{men}=0.591$, $p=0.031$). No significant differences between genders were found in Cyberoceptive and Interoceptive Errors.

\section{Data Analysis}
To more directly address our research questions, we conduct two types of correlation analysis, namely the analysis between (1) cyberoception and emotion and another between (2) cyberoception and interoception, 
\textbf{The analysis found a statistically significant correlation between the ``Turning On'' Cyberoception metric and emotional valence rating.} 

\subsection{Cyberoception and Emotion}
\textit{RQ1: Does cyberoception have similar \rev{emotion-related} properties to Interoception?}

\subsubsection{Correlational Analysis}
As mentioned in Section~\ref{sec:affectivepicture}, the numerical data of the participants' emotional experience was collected from the Affective Pictures Rating Experiment. 
The emotional experience data consists of two factors, literally valence and arousal. 
Pearson correlations between Cyberoception Metrics (Cyberoceptive Error) and two factors were analyzed. \rev{The unit of analysis is each participant.} The results are shown in Table~\ref{cybero_emotion}. 

Apparently, the Cyberoceptive Error of metrics is positively correlated to valence. \textbf{Especially, between the ``Turning On'' and valence, we found a correlation score of 0.452 (with statistical significance $p<0.052$).} 
It shows that \textbf{the participants who are more sensitive on ``Turnin On'' Cyberoception tend to experience emotions negatively}, even if the same stimulus is provided. 
To the best of our knowledge, we are the first to establish a correlation between the person's emotional valence value and their sense of smartphone usage.

Compared to valence, Cyberoception metrics showed a less significant correlation with arousal. We would like to further discuss in Section ~\ref{Discussion}. 


\aptLtoX[graphic=no,type=html]{\begin{table*}[t]
    \centering
    \caption{Correlation between Cyberoceptive Error and Emotion 
}
    \vspace{-0.2cm}
    \label{cybero_emotion}
    \begin{tabular}{c|ccccccc}
        \hline
        Valence & Turning On & Unlocking & Screen Use Duration & Micro-usage & Most-used App & Typo(daily-life) & Typo(in-lab)\\
        \hline\hline
        \begin{tabular}{c}$r$-value\\$p$-value\end{tabular} & \begin{tabular}{c}0.452\\ (0.052)\end{tabular} & \begin{tabular}{c}0.195\\ (0.425)\end{tabular} & \begin{tabular}{c}0.156\\ (0.488)\end{tabular} & \begin{tabular}{c}0.308\\ (0.187)\end{tabular} &\begin{tabular}{c}-0.220\\ (0.326)\end{tabular} & \begin{tabular}{c}0.253\\ (0.269)\end{tabular} & \begin{tabular}{c}-0.192\\ (0.446)\end{tabular}\\
        \hline
        \multicolumn{8}{c}{}\\
        \hline
        Arousal & Turning On & Unlocking & Screen Use Duration & Micro-usage & Most-used App & Typo(daily-life) & Typo(in-lab)\\
        \hline\hline
        \begin{tabular}{c}$r$-value\\$p$-value\end{tabular} & \begin{tabular}{c}0.149\\ (0.544)\end{tabular} & \begin{tabular}{c}0.056\\ (0.820)\end{tabular} & \begin{tabular}{c}-0.019\\ (0.934)\end{tabular} & \begin{tabular}{c}-0.096\\ (0.688)\end{tabular} & \begin{tabular}{c}0.041\\ (0.856)\end{tabular} &\begin{tabular}{c}0.113\\ (0.625)\end{tabular} & \begin{tabular}{c}-0.001\\ (0.997)\end{tabular}\\
        \hline
    \end{tabular}
\end{table*}}{\begin{table*}[t]
    \centering
    \caption{Correlation between Cyberoceptive Error and Emotion 
}
    \vspace{-0.2cm}
    \label{cybero_emotion}
    \begin{tabular}{c|ccccccc}
        \hline
        Valence & Turning On & Unlocking & Screen Use Duration & Micro-usage & Most-used App & Typo(daily-life) & Typo(in-lab)\\
        \hline\hline
        \begin{tabular}{c}$r$-value\\$p$-value\end{tabular} & \begin{tabular}{c}0.452\\\footnotesize(0.052)\end{tabular} & \begin{tabular}{c}0.195\\\footnotesize(0.425)\end{tabular} & \begin{tabular}{c}0.156\\\footnotesize(0.488)\end{tabular} & \begin{tabular}{c}0.308\\\footnotesize(0.187)\end{tabular} &\begin{tabular}{c}-0.220\\\footnotesize(0.326)\end{tabular} & \begin{tabular}{c}0.253\\\footnotesize(0.269)\end{tabular} & \begin{tabular}{c}-0.192\\\footnotesize(0.446)\end{tabular}\\
        \hline
        \multicolumn{8}{c}{}\\
        \hline
        Arousal & Turning On & Unlocking & Screen Use Duration & Micro-usage & Most-used App & Typo(daily-life) & Typo(in-lab)\\
        \hline\hline
        \begin{tabular}{c}$r$-value\\$p$-value\end{tabular} & \begin{tabular}{c}0.149\\\footnotesize(0.544)\end{tabular} & \begin{tabular}{c}0.056\\\footnotesize(0.820)\end{tabular} & \begin{tabular}{c}-0.019\\\footnotesize(0.934)\end{tabular} & \begin{tabular}{c}-0.096\\\footnotesize(0.688)\end{tabular} & \begin{tabular}{c}0.041\\\footnotesize(0.856)\end{tabular} &\begin{tabular}{c}0.113\\\footnotesize(0.625)\end{tabular} & \begin{tabular}{c}-0.001\\\footnotesize(0.997)\end{tabular}\\
        \hline
    \end{tabular}
\end{table*}}

\subsubsection{Regression Analysis}
\begin{figure}[t]
    \centering
    \includegraphics[width=\linewidth]{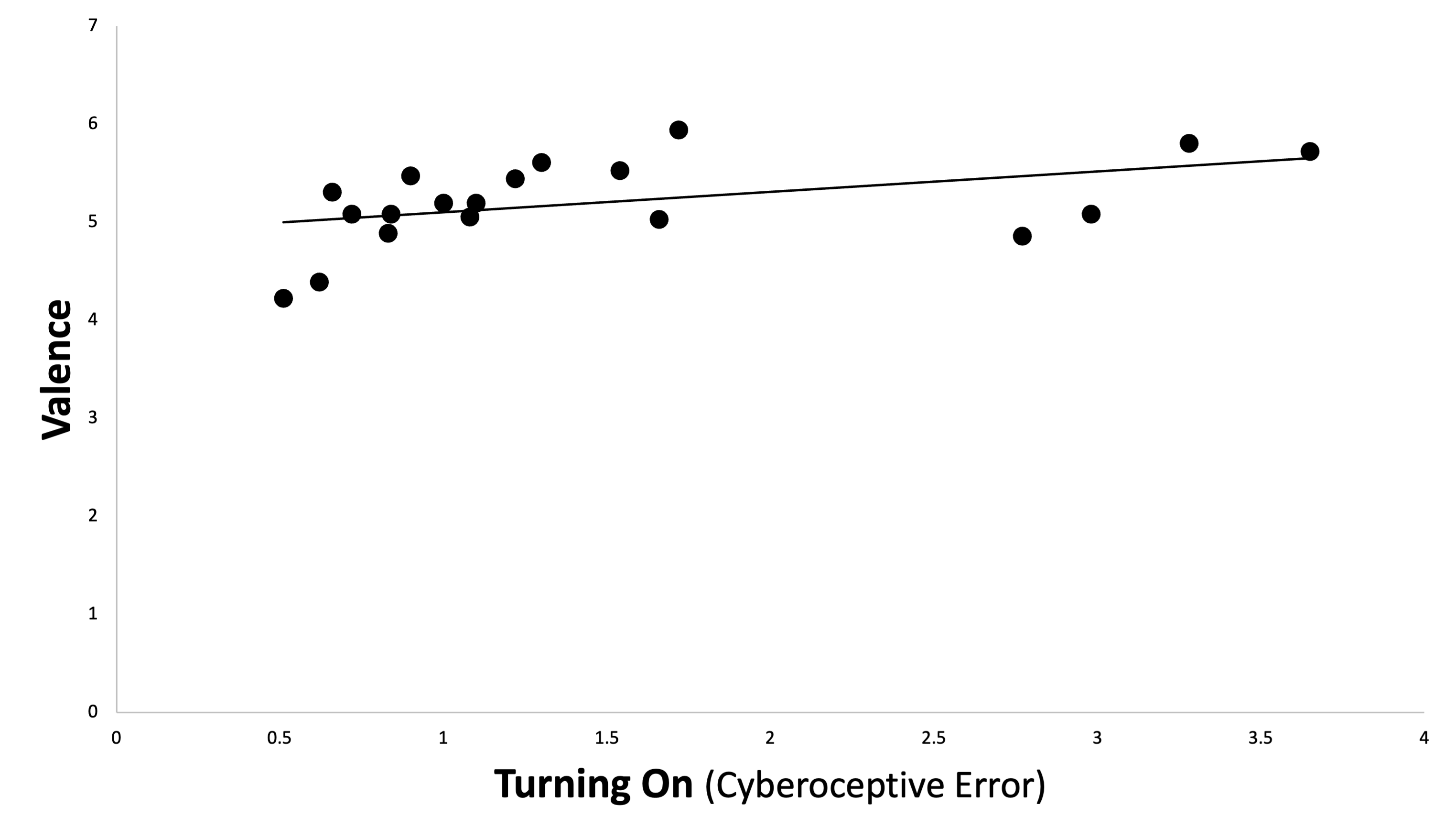}
    \caption{Correlation between Cyberoceptive Error of Turning On and Valence}
    \Description{Correlation between Cyberoceptive Error of Turning On and Valence}
    \label{TurningOn_Valence}
\end{figure}

A linear relationship between Cyberoception metrics and two emotional experience factors is established. \rev{The unit of analysis is each participant.} We found that ``Turning On - Valence'' tends to be significant($[\beta=0.208, t=2.089, p=0.052]$). The linear regression between Turning On and Valence is shown in Fig. \ref{TurningOn_Valence}. 

\subsubsection{Two-Split Analyses}
\begin{figure}[t]
    \centering
    \includegraphics[width=\linewidth]{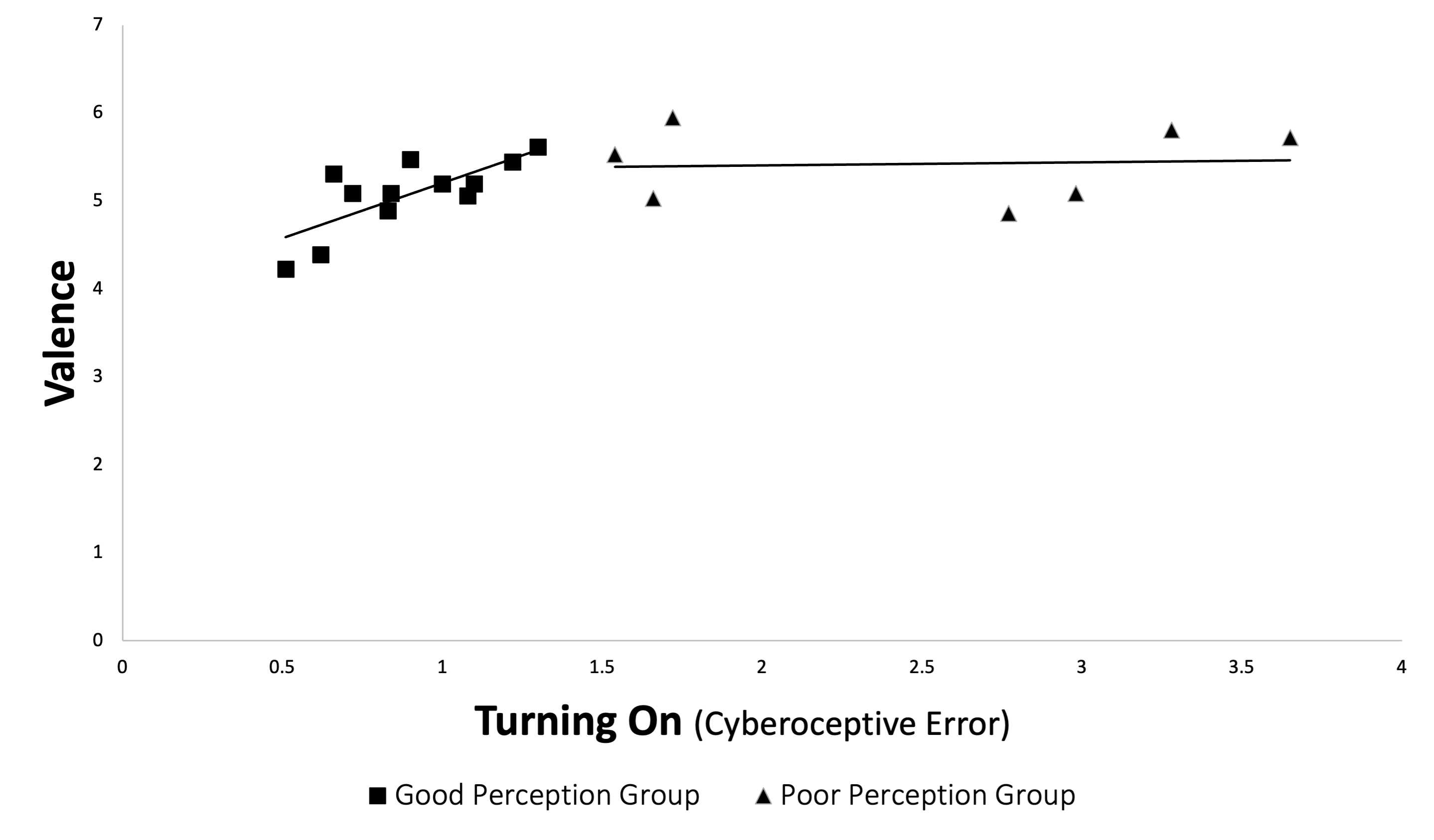}
    \caption{Two-Split Analyses on Cyberoceptive Error of Turning On and Valence}
    \Description{Two-Split Analyses on Cyberoceptive Error of Turning On and Valence}
    \label{Two-Split}
\end{figure}

We split participants into two groups using the average Cyberoceptive Error of their Turning On( M = 1.49 ). \rev{The unit of analysis is each participant. }
Among the Good Perception Group, the relationship between Turning On and Valence is significantly highly positive($[\beta=1.25, t=3.56, p=0.005]$). However, Among the Poor Perception Group, such a significant correlation was not observed($[\beta=0.03, t=0.15, p=0.88]$).

\subsubsection{Conclusion Regarding Research Question 1}
Our experimental result revealed that some of the cyberoception metrics we proposed (especially ``Turn On'' cyberoception) can work similarly to interoception, performing as an affective factor intimately connected to an individual's emotional experience. 
This correlation result between the cyberoception metric and valence supports the viewpoint that cyberoception has similar emotion-experience properties to interoception.

\subsection{Cyberoception and Interoception}

\textit{RQ2: Can cyberoception replace\rev{ the existing physiological sensor-based measurement methodology of} interoception?}

As shown in Table~\ref{tbl:cyberoception_interoception}, the intra-individual averaged error rate of interoception and cyberoception metrics are used to evaluate participants' Interoceptive Error and Cyberoceptive Error. 
Pearson correlation analyses were performed between Interoceptive Error and Cyberoceptive Errors.
Cyberoceptive errors of 2 types of Cyberoception Metrics (Turning On (0.154) and Micro-usage (0.255)) are positively correlated to the Interoceptive Error value.
Meanwhile, those of the other four types are negatively correlated to the Interoceptive Error value.


%

\subsubsection{Conclusion Regarding Research Question 2}
\rev{Regarding whether cyberoception can be replaced by interoception, we could not conclude that our proposed Cyberoceptive Metrics can replace interoception measurement. 
Though we could not confirm significant differences, for some of the proposed cyberoception metrics, we found weak correlations between the accuracy of cyberoception and that of interoception. These results show that the cyberoception for Turning On and micro-usage could be substituted to replace the interoception measurement.}

\aptLtoX[graphic=no,type=html]{\begin{table*}[t]
    \centering
    \caption{Correlation between Participants' Cyberoceptive Error and Interoceptive Error 
    }
    \vspace{-0.2cm}
    \label{tbl:cyberoception_interoception}
    \begin{tabular}{c|ccccccc}
        \hline
        Interoception & Turning On & Unlocking & Screen Use Duration & Micro-usage & Most-used App & \begin{tabular}{c}Typo\\(daily-life)\end{tabular} & \begin{tabular}{c}Typo\\(in-lab)\end{tabular}\\
        \hline\hline
        \begin{tabular}{c}$r$-value\\$p$-value\end{tabular} & \begin{tabular}{c}0.154\\ (0.528)\end{tabular} & \begin{tabular}{c}-0.264\\ (0.274)\end{tabular} &  \begin{tabular}{c}-0.170\\ (0.449)\end{tabular} & \begin{tabular}{c}0.255\\ (0.278)\end{tabular} & \begin{tabular}{c}0.060\\ (0.792)\end{tabular} &\begin{tabular}{c}-0.155\\ (0.504)\end{tabular} & \begin{tabular}{c}-0.248\\ (0.322)\end{tabular}\\
        \hline
    \end{tabular}
\end{table*}}{\begin{table*}[t]
    \centering
    \caption{Correlation between Participants' Cyberoceptive Error and Interoceptive Error 
    }
    \vspace{-0.2cm}
    \label{tbl:cyberoception_interoception}
    \begin{tabular}{c|ccccccc}
        \hline
        Interoception & Turning On & Unlocking & Screen Use Duration & Micro-usage & Most-used App & \begin{tabular}{c}Typo\\(daily-life)\end{tabular} & \begin{tabular}{c}Typo\\(in-lab)\end{tabular}\\
        \hline\hline
        \begin{tabular}{c}$r$-value\\$p$-value\end{tabular} & \begin{tabular}{c}0.154\\\footnotesize(0.528)\end{tabular} & \begin{tabular}{c}-0.264\\\footnotesize(0.274)\end{tabular} &  \begin{tabular}{c}-0.170\\\footnotesize(0.449)\end{tabular} & \begin{tabular}{c}0.255\\\footnotesize(0.278)\end{tabular} & \begin{tabular}{c}0.060\\\footnotesize(0.792)\end{tabular} &\begin{tabular}{c}-0.155\\\footnotesize(0.504)\end{tabular} & \begin{tabular}{c}-0.248\\\footnotesize(0.322)\end{tabular}\\
        \hline
    \end{tabular}
\end{table*}}

\subsection{Emotion and Operation Data}
We additionally analyzed all possible relations between six types of operation usage data obtained from smartphone API (not the ``sense'' of the operation but the operation itself) and the emotional and interoceptive data acquired in the in-lab experiment mentioned in \ref{sec:interoceptionexperiment} and \ref{sec:affectivepicture}. 

Interestingly, the ``Unlocking'' operation were correlated to arousal rating (r = 0.46, p = 0.05). Individuals who unlock their smartphones more frequently feel more excited or aroused than those who less frequent unlock. (For other combinations, we did not have statistically significant relations.)

\section{Discussion}
\label{Discussion}
This study introduces the concept of cyberoception, which can be an alternative to interoception as a factor of emotional ability. 
For the concrete operations representing Cyberoception, we proposed six types of Cyberoception metrics.
Through a ten-day daily-life study and in-lab emotional psychology experiment, we analyzed participants' perception of activities in cyberspace, specifically via their smartphones, interoceptive ability, and emotional individual characteristics.
\rev{As the answer of RQ1, we revealed that the ``Turn On'' cyberoception metrics can function similarly to interoception, performing as an affective factor that is intimately connected to an individual's emotional experience}
\rev{The result evidentiary supports the correlation between cyberoception and experience of emotional valence. }
However, \rev{As the answer of RQ2,} though cyberoception was hypothesized as a replacement for interoception \rev{measurement}, the experimental result rejected such a hypothesis and discriminated between cyberoception and interoception \rev{measurement}.

Although the experimental result did not support cyberoception to replace interoception \rev{measurement}, it is a valuable discovery that implies the unique characteristics that our proposed concept ``cyberoception'' holds.

One explanation for our result is that cyberoception, referring to the sense of the condition within cyberspace and parallels interoception, is crucial in users' personality traits on core affect and composes users' emotional experience. Future research explores the connection between cyberoception and varied emotions in users' daily lives, including long-term mood and short-term affective response. 

In previous literature~\cite {wiens, barrett2004interoceptive}, valence was demonstrated to be less related to interoception. 
Our findings of the correlation between the Cyberoceptive Error of Turning On and valence imply that cyberoception is connected to users' emotional experience through a different approach. Therefore, cyberoception can eliminate the blind point of interoception studies and provide a futuristic point of view toward users' affective states. 

Interoception is a unified sensation for each viscera, and it has been shown that visceral sensation to the stomach is consistent with that to the heart~\cite{herbert2012interoception}. In our study, we found a positive correlation among Cyberoception metrics. Although not statistically significant, such a positive correlation could imply a common nature among different Cyberoception metrics, which warrants further investigation.
Cyberoception should be positioned as a meta concept that includes multiple perceptions towards activities and conditions within cyberspace. 

Cyberoception employs usage data obtained from a user's daily smartphone activities, without any additional biometric sensors or well-controlled experimental environment. 
Thus, cyberoception enables real-time and non-invasive monitoring of user's emotional characteristics. Cyberoception, as a building block in other applications (such as a linked library), middleware, and even the operating system, contributes to understanding the affective context of users and providing more human-centered information services. 

\rev{In "measurement error in self-reported smartphone usage," previous studies have consistently highlighted discrepancies between self-reported smartphone usage and actual usage~\cite{lin2015time, douglas2012digital, felisoni2018cell}. While these studies have shown that measurement errors are non-random, the factors underlying these discrepancies remain underexplored. Most notably, demographic factors such as sex, age, marital status, and educational background have demonstrated limited or insignificant effects~\cite{shum2011evaluation,boase2013measuring}. By shifting the focus to affective states, this study uncovers a previously overlooked dimension, emphasizing the importance of emotional valence in shaping individuals' perceptions of their smartphone use.}

\subsection{Limitation and Future Work}
There are limitations in this study, which leaves room for further research.
In this study, we recruited participants from a university where many participants were underage (< 21). Therefore, the stimulus we showed them was strictly limited.
We avoided using erotic, violent, or grotesque pictures, which made the sampled pictures from IAPS tend to lean more toward a low-arousing distribution.

Since the participants were all university students, the inference of our experimental results is limited to the younger age group. Since it is common for interoception to decline with age~\cite{khalsa2009interoceptive}, our participants had better interoceptive ability than Schandry's study~\cite{schandry1981heart}. This biased age distribution can explain the difficulty encountered when investigating the relationship between interoception and other factors. Conducting experiments with all age groups would be essential for future work.

\rev{This study focuses only on the cardiac axes of interoception. Considering that much is still unknown about interoception, further research is required across other modalities, such as respiratory, gastric, colonic, and rectal axes.}

As it is challenging to fully understand emotions based solely on the physiological-based interoception, it is similarly difficult to comprehensively explain emotions independently through cyberoception.
The relationship between cyberoception and emotions demonstrated in this study represents only a partial aspect of the broader emotional experience. Further work is required in this area.

We conducted trials randomly orders within each in-lab experiment and forced adequate breaks between experiments. There is hardly any possibility of learning effects among the Affective Picture Rating Experiment, Interoception Experiment, and Typing Experiment because they are different tasks that follow distinctly different procedures. Future works could include investigating the effect of the mentioned in-lab experiments. 

\rev{One might argue that the method used in this experiment, which requires self-reporting every half hour, is not as non-invasive as measuring interoceptive sensation, which measures heart rate using a camera in daily or weekly intervals.
However, in the experiment in this paper, asking the subjects to self-report with such interval was one particular setting in the experiment to discover whether there was a correlation, given that it is still unknown how much each specific cyberoception metric correlates with emotional experience and interoception.
After cyberoception is discovered because of this research, it may be more effective to use a time interval other than 30 minutes for this sensation. Discovering a more effective time interval is outside the scope of this paper itself and can be considered in future.}

\rev{One notable limitation of this study, is the potential maladaptive consequences of heightened awareness of feelings. While this research aims to improve understanding of Cyberoception, psychology research suggests that consistently encouraging individuals to become acutely aware of their feelings can have unintended negative effects~\cite{nolen2008rethinking}. Furthermore, the coping techniques involving smartphone usage should be developed in future studies.}

\rev{In addition, there are potential risks associated with viewing smartphones as extensions of the body, a concept explored in this study. Excessive smartphone use is likely to intensify the phenomenon of the smartphone being perceived as an extension of the self, which may harm users' well-being. Given the widespread use of smartphones in contemporary society, there is an urgent need to accelerate research on the implications and dynamics of the ``smartphone as an extension of the self.'' In this context, research on cyberoception becomes crucial, and further studies are essential.}






\section{Conclusion}
In this study, we introduced the concept of cyberoception, which is expected to replace interoception as a factor of emotional ability.
Through a 10-day daily-life study and an in-lab emotional psychology experiment, we analyzed 22 participants' perceptions of activities in cyberspace, interoceptive ability, and individual emotional characteristics.
\rev{The result evidentiary support the correlation between the experience of emotional valence and a specific Cyberoception, ``Turning On''.}
Cyberoception can eliminate the blind point of interoception studies and provide a futuristic point of view toward users' affective states. We treat cyberoception as a key building block for many other apps that use the data to improve the health outcomes of mobile phone users.

\begin{acks}
This work was supported by JSPS KAKENHI Grant Number JP21K11853 and JP24K02935.
\end{acks}

\bibliographystyle{ACM-Reference-Format}
\bibliography{gao-cybero-CHI2024}

\appendix

\section{Questions in Daily Cyberoception Measuring Survey}
\label{appendix:questionsInEsmSurvey}
\begin{table}[H]
    \centering
    \caption{Questions in Daily Cyberoception Measuring Survey}
    \scalebox{0.85}{
    \begin{tabular}{c|c|c}
        \hline
        \textbf{Category} & \textbf{Likert Question} & \textbf{Numeric Question} \\
        \hline\hline
        Turning On & 
        \begin{tabular}[t]{p{0.5\linewidth}}
        Based on your daily experience with your smartphone use, during the period from 30 minutes ago to the present time, what do you think about your screen switching on while using this smartphone? Please answer based on your own perception/feeling and do not dwell on your answer.
        \end{tabular} & 
        \begin{tabular}[t]{p{0.2\linewidth}}
        Specifically, how many times did you switch the screen on during this 30min?
        \end{tabular}\\ 
        \hline
        Unlocking & 
        \begin{tabular}[t]{p{0.5\linewidth}}
        Based on your daily experience with your smartphone use, during the period from 30 minutes ago to the present time, what do you think about your smartphone unlocking while using this smartphone? Please answer based on your own perception/feeling and do not dwell on your answer.
        \end{tabular} &
        \begin{tabular}[t]{p{0.2\linewidth}}Specifically, how many times did you unlock this smartphone during this 30min?\end{tabular}
        \\
        \hline
        Screen Use Duration & 
        \begin{tabular}[t]{p{0.5\linewidth}}
        Based on your daily experience with your smartphone use, during the period from 30 minutes ago to the present time, what do you think about your screen usage duration? Please answer based on your own perception/feeling and do not dwell on your answer.
        \end{tabular} & 
        \begin{tabular}[t]{p{0.2\linewidth}}
        Specifically, how long did you use your screen during this 30min?
        \end{tabular}
        \\ 
        \hline
        Micro-usage & 
        \begin{tabular}[t]{p{0.5\linewidth}}
        Based on your daily experience with your smartphone use, during the period from 30 minutes ago to the present time, what do you think about your application usage under 5s, which means that you shut down an application in 5s from launching the application? Please answer based on your own perception/feeling and do not dwell on your answer.
        \end{tabular} & 
        \begin{tabular}[t]{p{0.2\linewidth}}Specifically, how many times did you shut down an application in 5s from launching the application?\end{tabular}
        \\
        \hline
        Most-used App & 
        \begin{tabular}[t]{p{0.5\linewidth}}
        Based on your daily experience with your smartphone use, during the period from 30 minutes ago to the present time, what do you think about your {Most Used Application Name} launching? Please answer based on your own perception/feeling and do not dwell on your answer.
        \end{tabular} & 
        \begin{tabular}[t]{p{0.2\linewidth}}Specifically, how many times did you launch this application?
        \end{tabular}
        \\ 
        \hline
        \begin{tabular}{c}Typo\\(daily-life)\end{tabular} & 
        \begin{tabular}[t]{p{0.5\linewidth}}
        Based on your daily experience with your smartphone use, during the period from 30 minutes ago to the present time, what do you think about the typo(Typographical Error) made by yourself on your smartphone? Please base on your own perception/feeling and do not dwell on your answer.
        \end{tabular} & 
        \begin{tabular}[t]{p{0.2\linewidth}}
        Specifically, how many times did you make a typo during this 30min?\end{tabular}\\
        \hline
    \end{tabular}
    }
\end{table}

\end{document}